\documentclass[12pt,preprint]{aastex}

\received{2004 April 19}
\begin{document}

\title{Measurements of Sunyaev-Zel'dovich Effect Scaling Relations for Clusters of
Galaxies}

\author{B.A. Benson\altaffilmark{1,2}, S.E. Church\altaffilmark{2}, P.A.R. Ade\altaffilmark{3}, J.J.
Bock\altaffilmark{4,5}, 
K.M.Ganga\altaffilmark{6}, C.N. Henson\altaffilmark{2,7}, 
\& K.L. Thompson\altaffilmark{2}} \altaffiltext{1}{Department of
Physics, 351 LeConte Hall, University of California, Berkeley, CA
94720} \email{bbenson@bolo.berkeley.edu} \altaffiltext{2}{Stanford
University, 382 Via Pueblo, Varian Building, Stanford, CA 94305}
\altaffiltext{3}{Department of Physics and Astronomy, University
of Wales, Cardiff, 5, The Parade, P.O. Box 913, Cardiff, CF24 3YB,
Wales, UK} \altaffiltext{4}{California Institute of Technology,
Observational Cosmology, M.S. 59--33, Pasadena, CA 91125}
\altaffiltext{5}{Jet Propulsion Laboratory, 4800 Oak Grove Dr.,
Pasadena, CA 91109} \altaffiltext{6}{Infrared Processing and
Analysis Center, MS 100-22, California Institute of Technology,
Pasadena, CA 91125} \altaffiltext{7}{Department of Physics,
University of California, 1 Shields Avenue, Davis, CA 95616}


\newcommand{\lan}{\langle}
\newcommand{\ran}{\rangle}
\newcommand{\be}{\begin{equation}}
\newcommand{\ee}{\end{equation}}
\newcommand{\raoff}{{\rm RA}_{\rm offset}}

\begin{abstract}
We present new measurements of the Sunyaev-Zel'dovich (SZ) effect
from clusters of galaxies using the Sunyaev-Zel'dovich Infrared
Experiment (SuZIE~II). We combine these new measurements with
previous cluster observations with the SuZIE instrument to form a
sample of 15 clusters of galaxies. For this sample we calculate
the central Comptonization, $y_0$, and the integrated SZ flux
decrement, $S$, for each of our clusters. We find that the
integrated SZ flux is a more robust observable derived from our
measurements than the central Comptonization due to inadequacies
in the spatial modelling of the intra-cluster gas with a standard
Beta model. This is highlighted by comparing our central
Comptonization results with values calculated from measurements
using the BIMA and OVRO interferometers. On average, the SuZIE
calculated central Comptonizations are $\sim60\%$ higher in the
cooling flow clusters than the interferometric values, compared to
only $\sim12\%$ higher in the non-cooling flow clusters. We
believe this discrepancy to be in large part due to the spatial
modelling of the intra-cluster gas. From our cluster sample we
construct $y_0$--$T$ and $S$--$T$ scaling relations. The
$y_0$--$T$ scaling relation is inconsistent with what we would
expect for self-similar clusters; however this result is
questionable because of the large systematic uncertainty in $y_0$.
The $S$--$T$ scaling relation has a slope and redshift evolution
consistent with what we expect for self-similar clusters with a
characteristic density that scales with the mean density of the
universe.  We rule out zero redshift evolution of the $S$--$T$
relation at $\sim90$\% confidence.

\end{abstract}

\section{Introduction}
\label{intro}

Clusters of galaxies are the largest gravitationally bound objects
in the Universe, and formed at relatively early times over a
critical redshift range ($0<z \lesssim 3$) during which the dark
energy came to dominate the total energy density of the Universe.
A measurement of the evolution of the cluster number density with
redshift is sensitive to various cosmological parameters,
including $\sigma_8$, $\Omega_M$, $\Omega_{\Lambda}$, and the dark
energy equation of state \citep{wang98,holder01}.  A direct
measurement of the cluster number density can be made through a
survey utilizing the Sunyaev-Zel'dovich (SZ) effect.  The SZ
effect is particularly well-suited for cluster surveys because an
SZ survey will detect every cluster above a mass limit that is
independent of redshift \citep[see][for example]{carl02}. Active
and planned SZ surveys should result in the discovery of tens of
thousands of clusters over the next several years, \citep[see][for
a review]{white03}.

The cosmological constraints from any SZ survey may ultimately be
limited by how closely clusters behave as standard candles.
\citet{haiman01} showed that future SZ surveys are likely to be
limited by systematic uncertainties due to the assumption that
clusters are virialized objects whose density scales with the mean
background density. Observations of relatively nearby clusters ($z
\lesssim 0.1$) in X-rays have shown that clusters are at least
remarkably regular objects whose observable properties seem to
obey well-behaved scaling relations. These scaling relations
include the mass-temperature \citep[e.g.,][]{fin01},
size-temperature \citep[e.g.,][]{mohr00}, and
luminosity-temperature \citep[e.g.,][]{mark98} scaling relations.
\citet{verde02} have argued that an integrated SZ flux versus
X-ray temperature scaling relation should have an exceptionally
small scatter, compared to other cluster scaling relations, and
should be especially useful in testing possible deviations from
virialization.

Investigations of SZ scaling relations have been limited so far
due to a scarcity of measurements. \citet{cooray99} and
\citet{mccarthyb} have compiled SZ measurements drawn from the
literature and constructed SZ scaling relations; however these
studies suffered from several draw-backs. Firstly, both papers
drew upon measurements from several different instruments. While
no systematic differences across instruments was known of at the
time, we believe this paper offers the first evidence that a
significant systematic discrepancy does exist. Secondly, both
papers concentrated on scaling relations that use the central
decrement of the cluster. The calculated central decrement often
relies on an assumed spatial distribution of the intra-cluster
(IC) gas, whose density is still best constrained by higher
resolution X-ray data. The traditional parameterization of the IC
gas distribution with the single Beta model
\citep{betaref1,betaref2} causes a large systematic uncertainty in
the central decrement calculated from SZ measurements. In this
paper, we address both these issues by constructing scaling
relations using SZ measurements from only one instrument, and
instead focusing on the integrated SZ flux scaling relation.


In this paper we describe new measurements of the SZ effect
towards eight galaxy clusters made with the Sunyaev-Zel'dovich
Infrared Experiment (SuZIE).  The SuZIE~II receiver makes
simultaneous measurements of the SZ effect in three frequency
bands, centered at 145\,GHz, 221\,GHz, and 355\,GHz, spanning the
null of the SZ thermal effect.  In combination with previous
measurements from the SuZIE instrument, we form a sample of 15
clusters for which we determine the central Comptonization, $y_0$,
and integrated SZ flux, $S$. We then use this sample to construct
scaling relations between $y_0$ and $S$ and the X-ray temperature
and compare these results to theoretical predictions. The layout
of this paper is as follows: in \S\ref{sec:obs}
and~\S\ref{sec:szanalysis} we describe the observations and
multi-frequency data analysis. In \S\ref{sec:145analysis} we
describe the analysis to determine the SZ flux. In
\S\ref{sec:y0compare} we compare our SZ measurements to other
experiments, particularly from the OVRO and BIMA interferometers.
Finally, in \S\ref{sec:scaling} we construct SZ scaling relations
and compare these measurements with theoretical predictions.

\section{S-Z Observations}
\label{sec:obs}
\subsection{Instrument}
We report measurements of the Sunyaev-Zeldovich effect made with
the second generation Sunyaev-Zeldovich Infrared Experiment
receiver (SuZIE~II) at the Caltech Submillimeter Observatory (CSO)
located on Mauna Kea.  The SuZIE~II receiver is a 2 $\times$ 2
array of three-color photometers that observes the sky
simultaneously in three frequency bands.  Radiation within each
frequency band is detected by a silicon nitride spider-web
bolometer cooled to 300 mK \citep{philbolo}.  The bolometer is
coupled to the primary mirror by a Winston horn which defines a
$\sim 1\farcm5$ FWHM beam on the sky, with each row separated by
$\sim 2\farcm3$ and each column by $\sim 5'$ on the sky. The
signal from the bolometers in the same row that are sensitive to
the same frequency are differenced electronically, equivalent to a
square wave chop of $5'$ on the sky.  For the new measurements
presented in this paper the three frequency bands in each
photometer were centered at 145, 221, and 355 GHz with
approximately 14\%, 11\%, and 9\% bandwidth respectively. For a
more detailed overview of the SuZIE~II beam sizes and pass-bands
see \citet{benson}.

\subsection{Observations}
In addition to the clusters reported in \citet{benson}, SuZIE~II
was used to make observations of an additional eight clusters over
the course of four observing runs during December 1998, January
2002, December 2002, and March 2003. These results are summarized
in Table~\ref{tab:obs}.  We selected bright, known X-ray clusters
from the {\em ROSAT} X-Ray Brightest Abell Clusters
\citep{1996MNRAS.281..799E, 1996MNRAS.283.1103E} and Brightest
Cluster Samples \citep{1998MNRAS.301..881E}.   In particular we
selected clusters with published intra-cluster (IC) gas density
models and electron temperatures that were previously unobserved
with SuZIE, or that had weak peculiar velocity constraints from
previous observations. We also observed clusters in directions on
the sky largely unobserved by SuZIE in order to have more uniform
sky coverage in our search for a local dipole flow
\citep[see][]{benson,church04}.

SuZIE~II operates in a drift scanning mode, where the telescope is
pointed ahead of the source and then parked.  The Earth's rotation
then causes the source to drift through the array of pixels.
Before each scan the dewar is rotated so that the rows of the
array lie along lines of constant declination.  Each scan lasts
two minutes, or $30'$ in right ascension, during which time the
telescope maintains a fixed altitude and azimuth.  After a scan is
complete, the telescope reacquires the source and the scan is then
repeated. Keeping the telescope fixed during an observation
prevents slow drifts from changes in ground-spillover from
contaminating the data.  From scan to scan the initial offset of
the telescope from the source is alternated between $12'$ and
$18'$, allowing a systematic check for an instrumental baseline
and a check for any time dependent signals. During the
observations presented here, the array was positioned so that one
row passed over the center of each cluster, as specified in
Table~\ref{tab:obs}.

\subsection{Calibration}
\label{sec:cal}

We observe planets for absolute calibration.  For the December
1998 and March 2003 observing runs Mars was used as the primary
calibrator, for the December 2002 run Uranus was used, and for the
January 2002 run Saturn was used.

We observe each primary calibrator at least once a night for
absolute calibration of the instrument. 
%
%
We correct for transmission of the atmosphere by measuring the
opacity using a 225 GHz tipping tau-meter located at the CSO.
During the March 2003 observing run this instrument was not
operational: for this run only, we used a scaled 225 GHz optical
opacity derived from the 350 $\mu$m tau-meter at the CSO.  The
opacity at 225 GHz is converted to the opacity in each of our
frequency bands by calculating a scaling factor $\alpha_{k}$ which
is measured from sky dips during stable atmospheric conditions.
Our responsivity to a celestial source is:
\be R = \frac{I_{\rm plan}\Omega_{\rm plan} \times e^{-\alpha
<\tau/\cos \theta_{\rm Cal}>}}{V_{peak}} \bigg(\frac{\rm Jy}{\rm
V}\bigg) \label{eqn:cal} \ee
where $\Omega_{\rm plan}$ is the angular size of the planet, and
$<\tau/\cos\theta_{\rm Cal}>$ is averaged over the length of the
observation, typically less than 20 minutes.

The brightness temperature of Uranus, Saturn, and Mars are all
well-studied at millimeter wavelengths.  Uranus has been measured
at millimeter wavelengths by \citet{griffin}, who model the
Uranian temperature spectrum, $T_{\rm Uranus}(\nu)$, with a third
order polynomial fit to the logarithm of wavelength.
\citeauthor{griffin} report a 6\% uncertainty in the brightness of
Uranus. \citet{goldin} report RJ temperatures of Saturn in four
frequency bands centered between 172 and 675 GHz. From these
measurements we fit a second order polynomial in frequency to
model $T_{\rm Saturn}(\nu)$.  \citeauthor{goldin} report a total
5\% uncertainty to the brightness of Saturn. We determine the
Martian temperature spectrum over our bands from the FLUXES
software package developed for the JCMT telescope on Mauna Kea.
\footnote{http://www.jach.hawaii.edu/JACpublic/JCMT/software/bin/fluxes.pl}
We fit the temperature spectrum given by FLUXES with a second
order polynomial fit to the logarithm of wavelength in order to
allow us to interpolate over our frequency range. The reported
uncertainty on the Martian brightness temperature is 5\%.

Generally it is preferable not to use Saturn as a primary
calibrator because of the unknown effect of its ring angle on its
millimeter wavelength emission.  However, previous SuZIE
measurements of Saturn at ring angles between $\pm 9^{\circ}$
showed a negligible effect on Saturn's millimeter wavelength
emission. During the December 2002 run calibration scans were
taken of Uranus and Saturn over several nights with Saturn at a
ring angle of $\sim -26.6^{\circ}$. Comparing the scans of the two
planets, Saturn was observed to have excess emission by $\sim 72,
43,$ and $37$\% in our 355, 221, and 145 GHz frequency bands.
Saturn was not used as a primary calibrator during this run;
however due to the lack of any other visible planets, Saturn was
the primary calibrator during the January 2002 run when Saturn was
at a ring angle of $\sim -25.8^{\circ}$. Because the ring angle of
Saturn changed by less than a degree between January 2002 and
December 2002, we use the cross-calibration of Saturn from Uranus
measured in December 2002 to correct the calibration from Saturn
for the January 2002 run. For the data presented in this paper,
this correction only affects the measurements of MS1054, see Table
\ref{tab:obs}.

Raster scans over Saturn and Mars were used to measure the beam
shape during each observing run.  In January and December 2002,
Saturn was used to measure the beam shape.  During these runs
Saturn had an angular diameter of $\sim 18.8''$ and $\sim 19.5''$
respectively. In December 1998 and March 2003, Mars was used to
measure the beam shape. During these runs Mars had an angular
diameter of $\sim 6.0''$ and $\sim 7.2''$, respectively. Compared
to our typical beam size of $\sim 90''$, these planets are
sufficiently small to be well approximated as point sources.


Including other sources of calibration uncertainty, such as from
beam uncertainties and detector non-linearities, we estimate the
total calibration uncertainty of SuZIE~II in each of its spectral
bands to be $\pm$10\%.  For a more thorough overview of the
SuZIE~II calibration uncertainty see \citet{benson}.



\section{Multi-frequency SuZIE Analysis and Results}
\label{sec:szanalysis}

Our analysis procedure is described in detail in \citet{benson},
but we briefly summarize it in this section. The basic philosophy
we use when analyzing SuZIE data is to perform a fit to the data
which uses the multi-frequency information to spectrally
distinguish between the SZ and atmospheric signal. This procedure
has proven invaluable to the analysis of our 355 GHz channel where
the atmosphere contributes the largest amount of extra noise.

\subsection{Defining the Data Set}

We will begin by defining some notation to aid in referencing the
SuZIE data set.  There are two rows of photometers in SuZIE.  Each
row consists of two photometers separated by $5'$ on the sky, with
a photometer consisting of three bolometers each observing at a
different frequency. In each row there are six single channel
signals, $S$, and three differenced signals, $D$.  The difference
signals are defined as the difference between the single channel
signals of the same frequency band in the same row.  We only
record three single channel signals from each row, one for each
frequency band. We use the subscript $k$ to refer to frequency
band, with $k=1,2,3$ referring to the frequency bands of 355, 221,
and 145 GHz in the on-source row.  We do not include analysis of
the data in the off-source row in this paper because it negligibly
improves the overall signal to noise. Each two minute scan is
broken into forty 3 second bins, with each bin covering a region
equal to $0\farcm75 \cos \delta$ on the sky. We will use the
subscripts $j$ to refer to scan number, and $i$ to refer to bin
number.  In this way the difference and single-channel signals at
145 GHz from scan $j$ and bin $i$ in the on-source row are
$D_{3ji}$ and $S_{3ji}$.

\subsection{SZ Model} \label{sec:szmodel}

The spatial model for the expected SZ signal in each scan is
calculated from the convolution of the beam map with a model of
the opacity of the cluster.  A beam map, $V_{k}(\theta, \phi)$, is
measured from raster scans of a planetary calibrator.  The optical
depth, $\tau$, of each cluster is modelled spatially by a
spherically symmetric isothermal $\beta$ model
\citep{betaref1,betaref2} with
\be \tau(\theta,\phi) = \tau_{0} \left(
1+\frac{\theta^2+\phi^2}{\theta_c^2}\right) ^{(1-3 \beta)/ 2} \ee
where $\theta$ and $\phi$ are angles on the sky and $\beta$ and
$\theta_c$ are parameters of the model. Because SuZIE is rotated
between scans to match the rotation of the sky, $\theta$ is
equivalent to right ascension and $\phi$ to declination.  For each
cluster the model parameters of the IC gas, $\beta$ and
$\theta_c$, are taken from the literature and are listed in Table
\ref{tab:icmodel}. In this table we also include the clusters
previously measured by SuZIE which will be included in our
analysis in section \ref{sec:145analysis}.  We define thermal and
kinematic SZ models, $S^{\rm th}(\theta)$ and $S^{\rm
kin}(\theta)$, to fit the data with
\be S^{\rm th}(\theta) = T_k m_k(\theta) \ee \be S^{\rm
kin}(\theta) = K_k m_k(\theta) \ee
where $m_k(\theta)$ is our spatial model, and $T_k$ and $K_k$ are
the thermal and kinematic band-averaged spectral factors which are
fully specified in \citet{benson}. The spatial model,
$m_k(\theta)$, is the convolution of the beam-shape, $V_{k}(\theta
',\phi)$, with the cluster optical depth, $\tau(\theta,\phi)$, and
has units of steradians. It is calculated at $0'.05\times cos
\delta$ intervals for a given offset, $\theta$, in right ascension
from the cluster center.  For each cluster the SZ spectral
factors, $T_k$ and $K_k$, are calculated assuming the X-ray
emission weighted temperature listed in Table \ref{tab:icmodel}.
With these definitions, $S^{\rm th}(\theta)$ and $S^{\rm
kin}(\theta)$ are the thermal and kinematic SZ signal we expect to
see in frequency band $k$ in a scan across a cluster of unity
central Comptonization, $y_0$, with a radial component to the
peculiar velocity, $v_{p}$, of 1\,km\,s$^{-1}$. The calculated SZ
model is then combined into $0'.75\times \cos \delta$ bins to
match the binned SuZIE~II data, so that we define $S^{\rm
th}_{ki}$ as the thermal SZ model in channel~$k$ for the right
ascension offset $\theta$ of bin number $i$.

\subsection{Atmospheric Noise Templates}

There are two sources of residual atmospheric noise in our
difference channels.  First there is a residual common signal to
each beam due to the finite common-mode rejection ratio (CMRR) of
the electronic differencing, and second there is a residual
differential atmospheric signal introduced because the two
differenced beams pass through slightly different columns of
atmosphere.  We model the residual common atmospheric signal,
$C_{kji}$, in each frequency band by the corresponding single
channel signal, $S_{kji}$.  We model the residual differential
atmospheric signal, $A_{ji}$, in each row by forming a linear
combination of the differential channels that contains no residual
SZ signal. We define $A_{ji} \equiv \alpha D_{1ji} + \gamma
D_{2ji} + D_{3ji}$, with the coefficients $\alpha$ and $\gamma$
chosen to minimize the residual thermal and kinematic SZ flux in
$A_{ji}$. In \citet{benson} the exact method to calculate $\alpha$
and $\gamma$ is explained in detail. We list the values of
$\alpha$ and $\gamma$ for each cluster in Table
\ref{tab:alphabeta}.  The values of $\alpha$ and $\gamma$ are
similar between clusters, but vary somewhat due to differences in
their intra-cluster gas model and electron temperature.

\subsection{Fitting for the Cluster Center} \label{sec:centerfit}

The location of the cluster in the scan is another free parameter
for which we can fit.  Before fitting for the cluster location we
perform an initial ``cleaning'' of each scan of each frequency
channel by removing the best-fit slope, constant offset, and a
signal proportional to the common and differential atmospheric
signal. The individual scans from each frequency channel are then
co-added, weighted by the root-mean squared of the residual of the
scan. The ``cleaned'' co-added scan from the on-source row at
$\nu\sim145$\,GHz ($k=3$) is then fitted with a model that
includes a slope, an offset, and the SZ model, where the cluster
location and the central Comptonization are allowed to vary.  The
best-fit cluster location, quantified as a right ascension offset,
$\raoff$, from the pointing center, is then used to fix the
cluster location for the rest of our analysis.  The values of
$\raoff$ for each cluster are listed in Table \ref{tab:finalfit}.
The nominal pointing center of each cluster is defined from its
respective X-ray centroid.

It is noted that from SuZIE data alone we can only constrain the
cluster's location in right ascension. The combined pointing
uncertainty from the CSO and the X-ray data is $\sim 20''$ and, in
general, is expected to have minimal effect on our results due to
the relatively large size, $\sim 90''$, of SuZIE's beams.  In fact
in \citet{benson} it was shown that a 20" pointing offset
corresponded to only a $\sim$4\% overestimate of the central
Comptonization, generally a negligible correction compared to the
statistical uncertainties in our measurements. For a more detailed
discussion of the pointing uncertainty and its effect on our
results see \citet{benson}.

The clusters observed in March 2003 all lie within 30" of the
nominal pointing center, however both clusters observed during
December 2002, A520 and A2390, are significantly off center. The
measured right ascension offset for A520 was $103^{+51}_{-53}$"
and for A2390 was $-60^{+28}_{-29}$". Because the offsets have a
different sign it is unlikely there was a systematic offset in our
pointing consistent across the sky.  In previous results, no
pointing error of this magnitude had been observed, of the eight
cluster observations reported in \citet{benson} all lie within 30"
of their nominal pointing centers. We now discuss the apparent
discrepancy of the cluster location of A520 and A2390
individually.

In Table \ref{tab:a520center} we give the location of A520 from
two different X-ray measurements and from our SuZIE observation.
The X-ray measurements are described in
\citet{1998MNRAS.301..881E} and \citet{allen00}, with the
respective locations calculated from the X-ray centroid.
No uncertainties were given for either X-ray measurement, however
the typical pointing uncertainty of ROSAT is $\sim 10''$ while the
spatial resolution of the PSPC is $\sim 30''$ and the HRI is $\sim
2''$. The pointing center for the SuZIE observation of A520 was
defined as the X-ray centroid of the PSPC.  It can be seen in
Table \ref{tab:a520center} that the X-ray centroid from the HRI is
more consistent with the location measured by SuZIE than the PSPC
observation which had defined our pointing center. In addition, an
observation of A520 by the OVRO interferometer measured a cluster
location consistent with the location measured by SuZIE
\citep{reesepc}.  This could suggest that the X-ray and SZ
centroid may not be equivalent for this cluster. For these
reasons, we believe the pointing offset observed by the SuZIE
observation of A520 is a real effect and consistent with the true
SZ center.

In Table \ref{tab:a520center} we give the location of A2390 based
on two different X-ray measurements and two different SuZIE
observations. The X-ray references for A2390 are the same as for
A520, and determine the cluster location from the position of the
X-ray centroid.  The X-ray centroid measured from the PSPC and
from the HRI agree very well.  Because the HRI has a much finer
spatial resolution than the PSPC, see the preceding paragraph, we
will only consider the more certain HRI coordinates. The SuZIE
observation from November 2000 measured a best-fit location nearly
coincident with the HRI X-ray centroid, while the SuZIE
observation from December 2002 measured a best-fit location
$\sim60$'' west. The 68\% confidence intervals do not overlap
between the measurements, however they are only separated by
$\sim9$''. Because we had not previously observed any pointing
offset and A2390 had been successfully observed with SuZIE before
in November 2000, we analyze the A2390 measurements from December
2002 assuming zero offset from the nominal pointing center.  This
adjustment changes the calculated central Comptonization of A2390
by only $\sim$3\% when considering the combined results of the
November 2000 and December 2002 observing runs which is calculated
in section \ref{sec:ssfits}.




\subsection{Individual Scan Fits for the Central Comptonization } \label{sec:ssfits}

To determine our best-fit results and formal confidence intervals
for the central Comptonization of each cluster, we have adopted
the approach of fitting all three frequency channels on a
scan-by-scan basis. From the raw binned calibrated data,
$D_{kji}$, we expect our signal to consist of a DC offset, a
slope, a signal proportional to our model of the residual common
and differential atmospheric signals, and a thermal and kinematic
SZ model, with a central Comptonization and peculiar velocity
fixed between frequency bands.  The residual signal, $R_{kji}$,
after removal of these signals is then
\be R_{kji} = D_{kji} - a_{kj} - ib_{kj} - (e_{kj} \times C_{kji})
- (f_{kj} \times A_{ji}) - y_{0j} T_k m_{k}(\theta_i-{\rm RA}_{\rm
offset}) - (y_0 v_{p})_{j} K_k m_{k}(\theta_i-{\rm RA}_{\rm
offset}) \label{eqn:sstvfits} \ee
where $a_{kj}$ are the offset terms, $b_{kj}$ are the slope terms,
and $e_{kj}$ ($f_{kj}$) are the coefficients that are proportional
to the common-mode (differential-mode) atmospheric signal in
frequency channels $k=1,2,3$. The SZ-model parameters $y_{0j}$ and
$(y_0 v_{p})_{j}$ are proportional to the magnitude of the thermal
and kinematic components in each frequency channel.  The best-fit
model of scan $j$ is then determined by minimizing the $\chi^2$ as
defined in \citet{benson}.  The uncertainty of the fit parameters
are determined using the standard definition from a general linear
least-squares fit \citep[see][for example]{press}.


From the individual scan fits we construct a 2 by 2 symmetric
covariance matrix, $\Sigma$, from each scan's fit for central
comptonization and peculiar velocity, see \citet{benson} for our
exact definition of $\Sigma$. Having calculated the covariance
matrix, we perform a maximum likelihood analysis for the model
parameters $y_{0}$ and $v_{p}$ with our likelihood function
defined as
\be L(v_{p},y_0) = \frac{1}{(2 \pi) |\mathbf{\Sigma}|^{1/2} } \exp
\bigg[ -\frac{1}{2} \left(
\begin{array}{c} <y_0>-y_0 \\ <y_0 v_{p}>-y_0 \times v_{p} \end{array} \right)^T \mathbf{\Sigma}^{-1}
\left(
\begin{array}{c} <y_0>-y_0 \\ <y_0 v_{p}>-y_0 \times v_{p} \end{array} \right)
 \bigg]
\ee
where $<y_0>$ and $<y_0 v_{p}>$ are weighted averages of the
individual scan fits for the thermal and kinematic SZ components,
and are defined as
\be <y_0> = \frac{\sum_j^{N_s} y_{0j}/\sigma^2_{y j}}{\sum_j^{N_s}
1/\sigma^2_{y j}} \label{eqn:ssy0} \ee \be <y_0 v_{p}> =
\frac{\sum_j^{N_s} (y_0 v_{p})_{j}/\sigma^2_{(y_0 v_{p})
j}}{\sum_j^{N_s} 1/\sigma^2_{(y v_{p}) j}} \label{eqn:ssy0vp} \ee
We calculate our formal confidence intervals for $y_0$ by
marginalizing $L(v_{p},y_0)$ over the peculiar velocity. See Table
\ref{tab:finalfit} for the $y_0$ results for each new cluster
presented in this paper.  The constraints on each cluster's
peculiar velocity will be the subject of a future paper.  In Table
\ref{tab:clustersummary} we give a summary of the $y_0$ results
calculated from a marginalization of $L(v_{p},y_0)$ for all the
clusters which have multi-frequency information from SuZIE.

\begin{figure}
\plotone{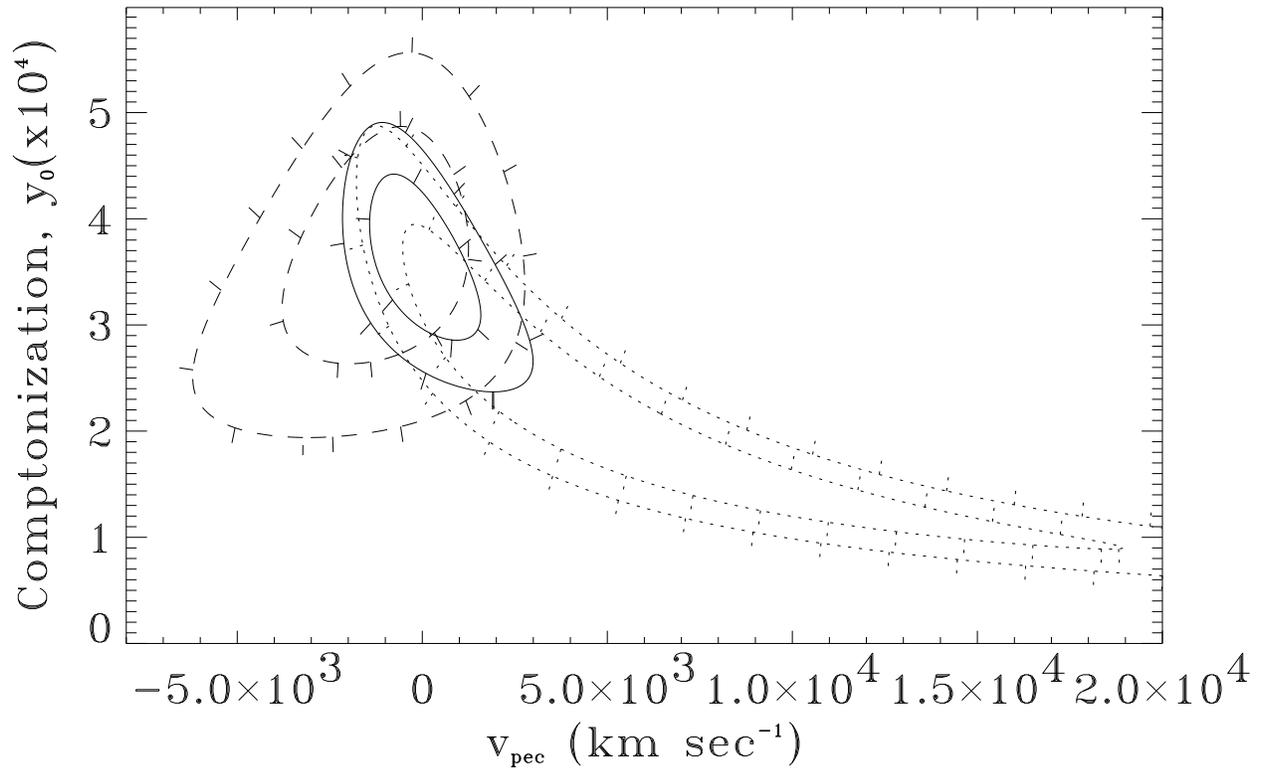} \caption[]{The two-dimensional likelihood of the
measurements of A2390 in November 2000, December 2002, and then
the combined likelihood from both observations. For each set of
data the 68.3\% and 95.4\% confidence regions are shown for peak
Comptonization and peculiar velocity.  The dotted contours are
from the November 2000 data, the dashed contours are from the
December 2002 data, and the solid contours are from the combined
likelihoods.} \label{fig:a2390like}
\end{figure}

Of the new measurements presented in this paper, only A2390 had
been observed previously by SuZIE. It is worthwhile to compare the
December 2002 results to the previous observation from November
2000 for a systematic check of any time-dependent or
observing-dependent errors. 
In Figure \ref{fig:a2390like} we plot the 2-d likelihoods from
both observing runs, and their product.  It is evident that the
overall constraints from the November 2000 data are much weaker.
Due to the low sensitivity of this data, there is a very weak
constraint on the peculiar velocity and a large degeneracy towards
an increasing peculiar velocity and a decreasing Comptonization.
The 68\% confidence regions do overlap between the two data sets,
and we consider them in good agreement. The combined likelihood
for A2390 is the product of the likelihoods from the November 2000
and December 2002 observing runs, $L(v_{p},y_0) = L(v_p,y_0)_{\rm
Nov00} \times L(v_p,y_0)_{\rm Dec02}$. For A2390, the value of
$y_0$ given in Table \ref{tab:clustersummary} is calculated from
marginalizing the combined likelihood function over peculiar
velocity.


\subsection{Spectral Plots for Each Cluster}
\label{sec:specplots}

In order to visualize the SZ spectrum measured by SuZIE we
calculate intensities from the co-additions of the differenced
data in each frequency band.  It should be noted, that the
calculated intensities and respective uncertainties given in this
section should not be used on their own due to the correlated
noise between frequency bands introduced by our atmospheric
subtraction method.  However, the best-fit intensities are an
accurate measurement of the correct intensity at each frequency,
and are meant to visually verify that we do measure an SZ-like
spectrum.

We begin with our cleaned data set, $Y_{kji}$, defined as
\be Y_{kji} = D_{kji} - a_{kj} - ib_{kj} - (e_{kj} \times C_{kji})
- (f_{kj} \times A_{ji}) \label{eqn:cleanedscan} \ee
with the best-fit parameters for $a_{kj}$, $b_{kj}$, $e_{kj}$, and
$f_{kj}$ determined from equation~(\ref{eqn:sstvfits}).  This
cleaned data set can now be co-added using the RMS of $R_{kji}$, from
equation \ref{eqn:sstvfits}, as a weight, such that:
\be Y_{ki} = \frac{\sum_{j=1}^{N_s} Y_{kji}/{\rm RMS}^2_{kj}}
{\sum_{j=1}^{N_s} 1/{\rm RMS}^2_{kj}} \label{eqn:cleanedcoadd}\ee
with ${\rm RMS}^2_{kj} \equiv \sum_i^{N_b} (R_{kji})^2/(N_b-1)$.
The RMS of $R_{kji}$ is not biased by any contribution from the SZ
source.  The uncertainty of each co-added bin, $\sigma_{ki}$, is
determined from the dispersion about the mean value, $Y_{ki}$,
weighted by RMS$_{kj}^2$,
\be \sigma_{ki} =
\sqrt{\frac{\sum_{j=1}^{N_s}\,(Y_{ki}-Y_{kji})^2/{\rm RMS}^2_{kj}}
{(N_s-1)\sum_{j=1}^{N_s} 1/{\rm RMS}^2_{kj}}} \ee
The best-fit central intensity, $I_k$, for each frequency band is
then found by minimizing the $\chi_k^2$ of the fit to the co-added
data, where $\chi_k^2$ is defined as follows:
\be \chi_k^2 = \sum_{i=1}^{N_b} \frac{\left[ Y_{ki}-I_k \times
m_{k}(\theta_i-{\rm RA}_{\rm offset}) \right]^2 }{ \sigma_{ki}^2 }
\label{eqn:chispec} \ee
We calculate the confidence intervals for $I_k$ under the
assumption of Gaussian errors on $Y_{ki}$ where our likelihood
function is related to the $\chi^2$ via $L(I_k) \propto \exp
(-\chi_k^2/2)$.

\begin{figure} \epsscale{0.75}
\plotone{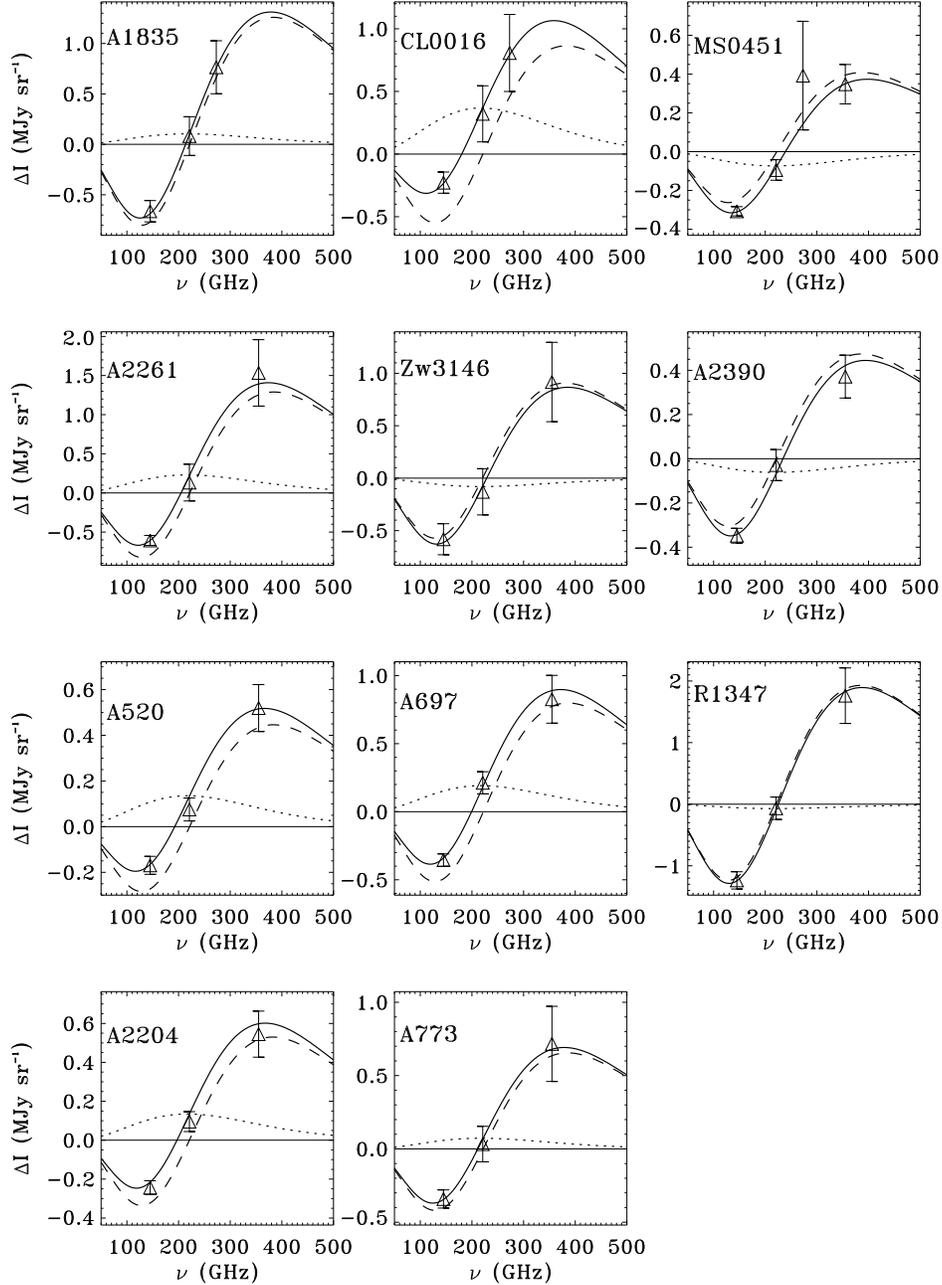} \caption[]{The measured SZ spectrum for each
cluster observation reported in this paper and \citet{benson}. In
each plot the solid line is the best-fit SZ model, the dashed line
is the thermal component of the SZ effect and the dotted line is
the kinematic component of the SZ effect.}
\label{fig:clusterspectra}
\end{figure}

Figure \ref{fig:clusterspectra} plots the best-fit SZ spectrum
with the measured intensities overlaid for the entire sample of 11
clusters detected in multiple frequency bands by SuZIE~II. This
plot includes cluster spectra given in \citet{benson}, along with
the new results from clusters observed in the December 2002 and
March 2003 observing runs.  For clusters detected in multiple
observing runs, such as for MS0451 and A2390, the intensity points
are calculated by multiplying the likelihoods in intensity space,
$L(I_k)$, from each observing run for each respective frequency
band.

\section{145 GHz SuZIE Analysis and Results} \label{sec:145analysis}

Ultimately we wish to compare our results for the central
Comptonization to independent single frequency SZ measurements,
which by themselves cannot constrain the peculiar velocity. In
principle, our multi-frequency results should be an appropriate
comparison because they take full account of the shape of the SZ
spectrum and should therefore accurately measure Comptonization.
However, there are some advantages in considering the 145 GHz data
on its own.  The 145 GHz channel is the most sensitive of the
frequency channels to the SZ thermal effect.  Including the higher
frequency channels allows one to constrain the peculiar velocity
as well as the central Comptonization.  However, due to the lower
sensitivity of the higher frequency channels and the addition of
the peculiar velocity as a free paramater, the overall constraints
on the calculated central Comptonization actually decreases with
the addition of the higher frequency channels. In addition, the
higher frequency channels suffer more confusion from
sub-millimeter point sources.  We showed in \citet{benson} that
typical sub-millimeter point sources in our cluster fields have a
tendency to bias our peculiar velocity results towards negative
values by a factor of several hundred kilometers per second and
our central Comptonization results higher by several times
$10^{-4}$.  This effect can be minimized by only analyzing the 145
GHz data because sub-millimeter point sources have spectral energy
densities which decrease with frequency. For these reasons, it is
useful to analyze the cleaned co-added 145 GHz data alone, which
will be the subject of this section.  For this analysis we include
previous SuZIE measurements given in \citet{swlh97b} and
\citet{benson} as well as the measurements given in this paper.



\subsection{Fitting for a Central Comptonization} \label{sec:145y0}

We calculate a central Comptonization from the co-added 145 GHz
data using a method similar to the one used to calculate the
intensity points in section \ref{sec:specplots} except here we are
solving for a central Comptonization instead.  For our data, we
use the co-added scan from the on-source row at 145 GHz, $Y_{3i}$,
which was defined in equation \ref{eqn:cleanedcoadd}. From
$Y_{3i}$ we subtract a SZ model which includes a peculiar
velocity, $v_p$, and a central Comptonization, $y_0$, to calculate
a $\chi^2$ which we define as
\be \chi^2(v_p,y_0) = \sum_{i=1}^{N_b} \frac{\left[ Y_{3i}- y_0
m_{3}(\theta_i-{\rm RA}_{\rm offset})(T_3 + v_p K_3) \right]^2 }{
\sigma_{3i}^2 } \label{eqn:chiy0} \ee
where $T_3$, $K_3$, and $m_3(\theta)$ are defined in section
\ref{sec:szmodel}.  Under the assumption of Gaussian errors on
$Y_{3i}$, this $\chi^2$ is related to the likelihood via
$L(v_p,y_0) \propto \exp(-\chi^2(v_p,y_0)/2)$. In this work we are
interested only in the central Comptonization, and so marginalize
over the peculiar velocity.  We assume a Gaussian prior on $v_p$
whose likelihood we take to be $\propto \exp(-v_p^2/2\sigma_v^2)$.
For these measurements we assume a most-likely peculiar velocity
of $v_{p}=0$ km s$^{-1}$ and a Gaussian width $\sigma_v=500$ km
s$^{-1}$. Because the clusters peculiar velocities are expected to
be randomly distributed around $v_{p}=0$ km s$^{-1}$, this
assumption should not bias these results.  We marginalize over
peculiar velocity such that our formal probability distribution
for the central Comptonization, $P(y_0)$, is defined as
\be P(y_0) = \frac{1}{\sqrt{2 \pi \sigma_v^2}} \int L(v_p,y_0)
\exp\left(\frac{-v_p^2}{2\sigma_v^2}\right) dv_p \label{eqn:ly0}
\ee
From $P(y_0)$ we calculate our best-fit central Comptonization and
associated 68\% confidence region for the 15 clusters presented in
this paper, \citet{benson}, and \citet{swlh97b}.  For a summary of
these results see Table \ref{tab:d3y0}.

With currently favored cosmological models it is expected that the
peculiar velocities of clusters be less than 1000\,km\,s$^{-1}$
\citep{gramann95, sheth01, suhh02}. However recent observations
show evidence for internal flows as large as 4000\,km\,s$^{-1}$
\citep{dupke02, mark03}.  Therefore it is expected that the
$\sigma_v=500$\,km\,s$^{-1}$ prior is a reasonable estimate of the
true width, however larger velocities have not been
observationally ruled out. Because the multi-frequency results
from SuZIE constrain the peculiar velocity through the measurement
of the SZ spectrum, we will consider a broader range of priors on
the peculiar velocity when we re-analyze these results in section
\ref{sec:suzieovrocompare}. There we will show that broadening
this prior to include larger peculiar velocities with greater
probability does not greatly affect the results.  However for the
145 GHz data analysis we only consider the case where
$\sigma_v=500$\,km\,s$^{-1}$.

If we compare the central Comptonizations calculated from the 145
GHz data, given in Table \ref{tab:d3y0}, to those calculated from
the multi-frequency data, given in Table \ref{tab:clustersummary},
it is clear that the 145 GHz results give better constraints for
the central Comptonization.  While it may seem counterintuitive
that the exclusion of two frequency channels actually increases
the constraints on the central Comptonization, this gain occurs
because of the way we handle the cluster peculiar velocity in both
calculations.  For the 145 GHz analysis we placed a Gaussian prior
of width 500 km s$^{-1}$ on the peculiar velocity.  However, for
the multi-frequency analysis in section \ref{sec:ssfits}, we
placed no prior on the peculiar velocity, adding a degree of
freedom to the analysis.  In fact, the multi-frequency constraints
on peculiar velocity are $\sim1000-2000$ km s$^{-1}$
\citep{benson,church04}, which is less constraining than the prior
we used in the 145 GHz analysis.  Because the higher frequency
channels are also less sensitive to the SZ thermal effect than the
145 GHz channel, the overall effect is that the 145 GHz results
more tightly constrain the central Comptonization than the
multi-frequency results.


It should be noted that for the clusters from \citet{swlh97b},
A1689 and A2163, the IC gas model used by \citet{swlh97b} differed
from those used in \citet{reese02}.  In Table \ref{tab:suziei} we
give a summary of the beta models and electron temperatures used
in both references.  We also give the calculated central
Comptonization derived from the SuZIE measurements using the two
sets of IC gas models.  For the case of A1689 the difference in
beta model parameters was significant. This is not surprising
considering the gas model used by \citet{swlh97b} for A1689 was
calculated from an unpublished analysis of a PSPC observation,
while the model used by \citet{reese02} was calculated from a more
recent HRI observation.  For the A1689 SuZIE results, the model
assumed significantly changes the calculated central
Comptonization, by $\sim40\%$. This difference is largely because
A1689 is unresolved by SuZIE, and therefore the central
Comptonization calculated depends entirely on the assumed IC gas
model.  To maintain consistency with \citet{reese02}, the central
Comptonizations of A1689 and A2163 given in Table \ref{tab:d3y0}
assume the IC gas model parameters used in \citet{reese02}.
Because all of the clusters observed by SuZIE are unresolved the
calculated central Comptonization depends sensitively on the
assumed IC gas model, this uncertainty will be discussed further
in section \ref{sec:systematics}.

\subsection{Fitting for the Integrated SZ Flux} \label{sec:szint}

In the last section, it was shown that the inferred central
Comptonization for A1689 changes significantly depending on the IC
gas model assumed.  It would be preferable to express our SZ
measurements in a way that depends less sensitively on the assumed
IC gas model.  An alternative observable is the SZ flux integrated
over some well-defined area on the sky.
In the literature, this area is usually defined by the radius at
which the mean over-density of the cluster is equal to some
factor, $\Delta$, times the critical density of the universe at
that redshift, $\rho_{\rm clust}(r_{\Delta}) = \rho_{\rm
crit}(r_{\Delta}) \Delta$. For X-ray measurements the value of
$\Delta$ is usually chosen in a range between 500 and 2500 because
the intra-cluster gas is expected to be virialized within this
range of radii \citep{evrard96}.  In this paper we adopt
$\Delta=2500$ with the $r_{2500}$ calculated for each cluster
given in Table \ref{tab:cluststuff}. For this choice of $\Delta$,
$r_{2500}$ is less than the SuZIE 5' difference chop, assuming a
standard $\Lambda$CDM cosmology, for each cluster. In this section
we will detail our calculation of the integrated SZ flux within
$r_{2500}$, $S(r_{2500})$.

The total mass of a cluster whose gas distribution is described by
an isothermal $\beta$ model can be calculated, under the
assumption of hydrostatic equilibrium and spherical symmetry, such
that the total mass within a radius, $r$, is
\be M_{\rm clust}(r) = \frac{3 k T_e \beta}{G \mu m_p}
\frac{r^3}{r_c^2+r^2} \ee
where $T_e$ is the cluster's electron temperature, $\mu m_p$ is
the mean molecular weight of the gas where we assume $\mu=0.6$,
with $\beta$ and $r_c$ corresponding to the $\beta$ model
parameters for the cluster. The cluster mass can be related to the
critical density of the universe, $\rho_{\rm crit}$, by
\be M_{\rm clust}(r_{\Delta}) = \rho_{\rm crit}(z) \frac{4 \pi
r_{\Delta}^3}{3} \Delta \label{eqn:clustmass} \ee
where $\rho_{\rm crit}(z)=3H(z)^2/(8\pi G)$, $H(z)$ is the Hubble
constant at a redshift $z$, $G$ is the Gravitational constant,
$r_{\Delta}$ is some radius of the cluster, and $\Delta$ is the
constant which makes this expression true.  For reasons given at
the beginning of this section we adopt $\Delta=2500$.  Equation
\ref{eqn:clustmass} can be re-arranged to solve for $r_{2500}$
with
%
%
%
\be r_{2500} = \left[ \frac{6}{2500} \frac{kT_e \beta}{\mu m_p}
\frac{c^2}{H_0^2 E(z)^2} - r_c^2 \right]^{1/2} \label{eqn:r2500}
\ee
where the variables are previously defined and where we have
replaced $H(z)^2 = E(z)^2 H_0^2$ with $E(z)^2\equiv \Omega_M
(1+z)^3 + (1-\Omega_M-\Omega_{\Lambda})(1+z)^2 +
\Omega_{\Lambda}$. We can then define the integrated flux as
\be S(r_{2500}) = y_0 T_k \int^{r_{2500}/d_A}_0 2\pi\theta
\left(1+\frac{\theta^2}{\theta_c^2}\right)^{(1-3\beta)/2} d\theta
\label{eqn:szint} \ee
where $y_0$ is the central Comptonization, $T_k$ is the thermal SZ
band-averaged spectral factor used in equation \ref{eqn:sstvfits}
and fully specified in \citet{benson}, $\theta_c$ and $\beta$ are
the IC gas model parameters, and $d_A$ is the angular diameter
distance to the cluster.  We note that $T_k$ depends on the
cluster electron temperature due to relativistic corrections to
the SZ spectrum. For the current SuZIE~II 145 GHz (k=3) bandpass
$T_3 = A \times 2(kT_{\rm CMB})^3/(hc)^2$ where $A=-3.93$ in the
non-relativistic limit and varies between $-3.6$ and $-3.8$ over
our typical range of electron temperatures.

We assume the following in our calculation of $S(r_{2500})$ from
equation \ref{eqn:szint}.  For all the clusters in our sample, we
have assumed the 145 GHz (k=3) band for the current SuZIE
instrument, and we use the central Comptonization results which
were calculated in section \ref{sec:145y0}, whose values are given
in Table \ref{tab:d3y0}.  For all clusters we assume the IC gas
model parameters given in Table \ref{tab:icmodel}.  More
precisely, for the non-cooling flow clusters we use the X-ray
emission weighted temperatures, and for the cooling flow clusters
we use the X-ray temperatures which account for the presence of
the cooling flows. It is well-known from X-ray measurements, that
cooling flows bias X-ray measured temperatures low compared to the
virial temperature of the IC gas \citep[see][for example]{af98a}.
The central Comptonization results given in section
\ref{sec:145y0} assume the standard X-ray emission weighted
temperature, even for the cooling flow clusters, however we will
show in section \ref{sec:teerror} that this correction is
negligible compared to the statistical uncertainty in our results.
Making the above assumptions we calculate $S(r_{2500})$ for each
cluster, these results are given in Table \ref{tab:cluststuff}.

The error bars for $S(r_{2500})$ given in Table
\ref{tab:cluststuff} are calculated from the statistical
uncertainty in $y_0$ and $T_e$ added in quadrature according to
equation \ref{eqn:szint}.  In general, the statistical uncertainty
in $y_0$ dominates the total uncertainty in $S(r_{2500})$. For
example, in the case of RXJ1347, the overall uncertainty in
$S(r_{2500})$ is $\sim35$mJy with the temperature uncertainty
contributing an uncertainty of $\sim8$mJy in $S(r_{2500})$, which
when added in quadrature is negligible. However, several clusters
have significantly less constrained electron temperatures,
particularly those clusters with cooling flow corrected
temperatures based on ASCA data, for which the temperature
uncertainty is a significant contribution to the overall
uncertainty in $S(r_{2500})$.


\subsection{Systematic Effects from an Uncertain Beta Model}
\label{sec:systic}

In order to examine the effect of an uncertain Beta model on our
145 GHz results, we have calculated the central Comptonization,
and the integrated SZ flux out to $r_{2500}$ for a range of Beta
models for two of our clusters, A1689 and A1835. We have chosen
these clusters because they are cooling flow clusters, whose X-ray
emission, in general, is not as well characterized by a spherical
isothermal Beta model.  In addition, both clusters have at least
two published Beta gas models derived from different X-ray
instruments sensitive to different spatial scales.  One of the
systematic effects that we are primarily concerned about in using
X-ray models to fit SZ observations is that the X-ray Beta models
might over-fit to the cooling core for cooling flow clusters. This
is a result of X-ray observations being more sensitive to
over-densities in the core of a cluster than SZ observations,
because $L_X \propto n_e^2$ while $I_{SZ} \propto n_e$.  By
choosing models derived from X-ray data which either do not
resolve the cooling core, or exclude it entirely in the spatial
fit, we can get some idea of the range of derivable Beta models
from X-ray data for both clusters.

\begin{figure} \epsscale{1}
\plotone{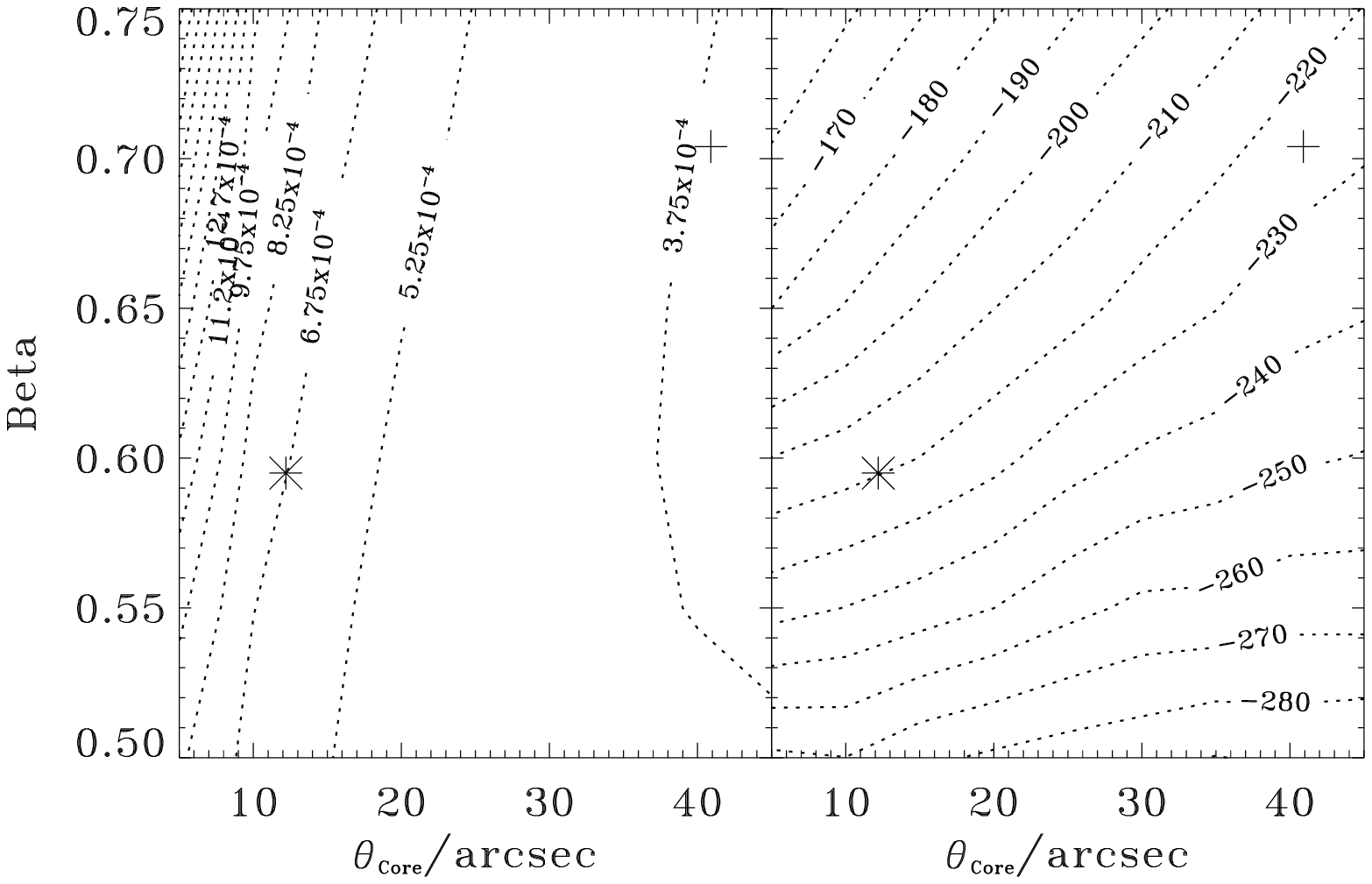} \caption[]{The best-fit $y_0$ and $S(r_{2500})$
calculated for A1835 using a range of Beta models. The asterisk
marks where the location of the gas model given in
\citet{reese02}, and the plus sign marks the location of the Beta
model given in \citet{majer02}, which fits only the outer region
of the cluster. Left: The central Comptonziation of A1835, the
contour levels spaced in $1.5\times10^{-4}$ intervals. Right: The
integrated SZ flux at 145 GHz, $S(r_{2500})$, from A1835, the
contour levels are spaced in 10 mJy intervals.} \label{fig:a1835}
\end{figure}

Figure \ref{fig:a1835} shows the calculated central
Comptonization, $y_0$, and integrated SZ flux, $S(r_{2500})$, for
A1835, assuming a suitable range of Beta models. In the figure the
asterisk denotes the Beta model given in \citet{reese02}, derived
from ROSAT-HRI data, and the plus sign denotes the Beta model
given in \citet{majer02}, which was derived from observations with
XMM and excluded the central region of the cluster out to a radius
of 42 arcsec.  The choice of Beta model causes nearly a factor of
2 difference in the calculated central Comptonization between the
two IC gas models. However, $S(r_{2500})$ varies by $\lesssim 3\%$
between the same models, with the line which connects the two
models nearly lying along a line of constant integrated flux.

\begin{figure}
\plotone{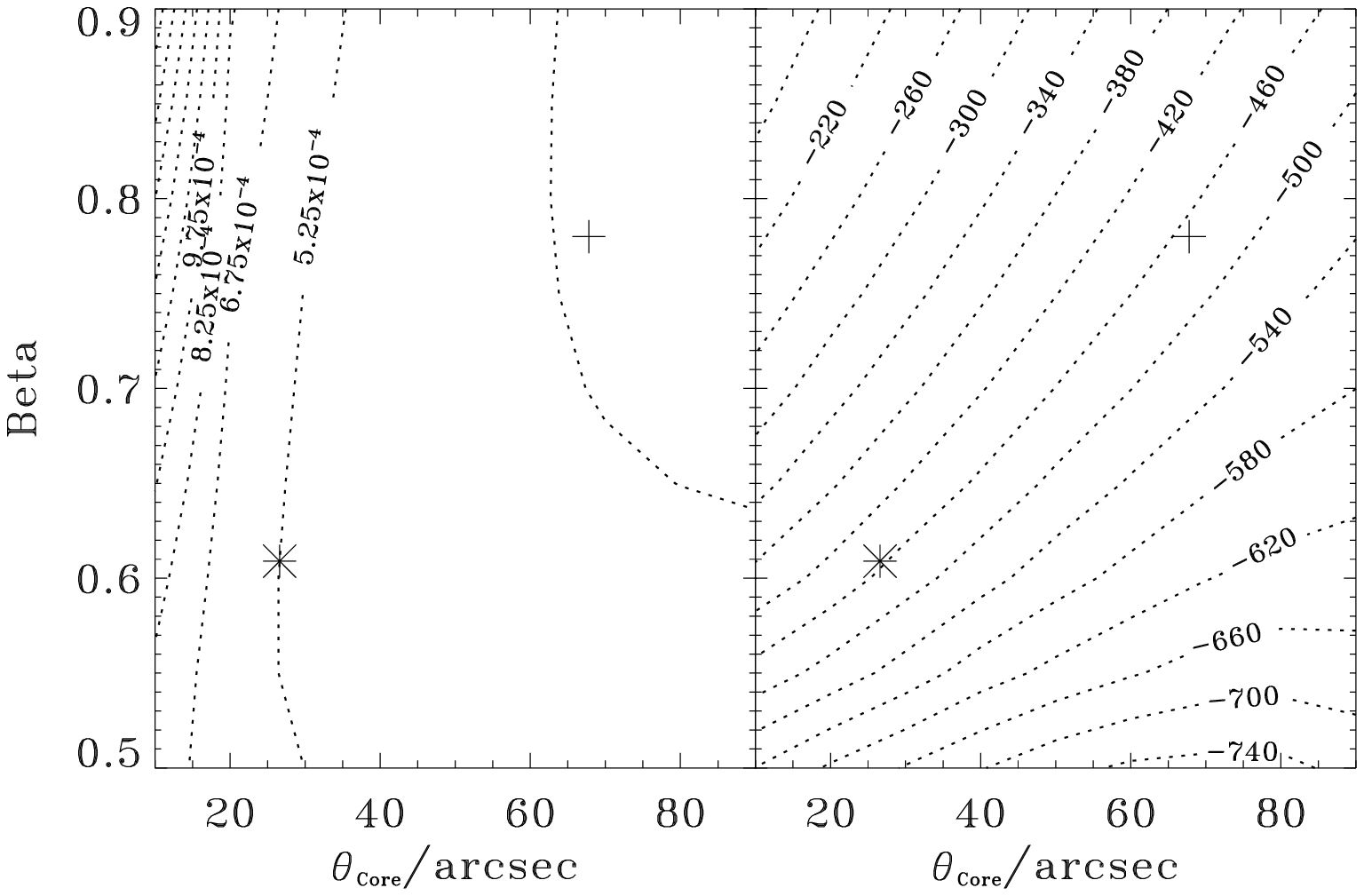} \caption[]{The best-fit $y_0$ and $S(r_{2500})$
calculated for A1689 using a range of Beta models. The asterisk
marks where the location of the gas model given in
\citet{reese02}, and the plus sign marks the location of the Beta
model given in \citet{swlh97b}. Left: The central Comptonziation
of A1689, the contour levels spaced in $1.5\times10^{-4}$
intervals. Right: The integrated SZ flux at 145 GHz,
$S(r_{2500})$, from A1689, the contour levels are spaced in 40 mJy
intervals.} \label{fig:a1689}
\end{figure}

Figure \ref{fig:a1689} shows the calculated central
Comptonization, $y_0$, and integrated SZ flux, $S(r_{2500})$, for
A1689, assuming a suitable range of Beta models. In the figure the
asterisk denotes the Beta model given in \citet{reese02}, derived
from ROSAT-HRI data, and the plus sign denotes the Beta model
given in \citet{swlh97b}, derived from ROSAT-PSPC data. Again,
which IC gas model is assumed significantly changes the
calculated central Comptonization, by $\sim 40\%$, however between
the two models $S(r_{2500})$ varies by $\lesssim 2\%$.

As we have seen, for both A1835 and A1689 the calculated central
Comptonization is much more sensitive to the assumed Beta model
than is the integrated SZ flux.  The physical reason why this is
true is because both clusters are unresolved by SuZIE and
therefore the calculated central Comptonization depends entirely
on the assumed IC gas model. Conversely, $S(r_{2500})$ is less
sensitive to the assumed IC gas model because $r_{2500}$ is
well-matched to the SuZIE beam-size, with $r_{2500}$ within a
factor of $\sim2$ of the SuZIE beam-size for all our clusters. For
our two example clusters, A1835 and A1689, $S(r_{2500})$ varies by
$\lesssim 3\%$ even when significantly different Beta models
derived from X-ray measurements with different spatial resolutions
are used. The results for A1835 and A1689 imply that even if the
X-ray IC gas model over-fits to the cooling core, this does not
have a significant systematic effect on the value of $S(r_{2500})$
derived from the SuZIE measurements. We therefore conclude that
the choice of Beta model adds a negligible uncertainty to
$S(r_{2500})$ when compared to the statistical uncertainty of our
data.

We should note that extending our integrated flux calculations to
larger radii potentially increases the systematic error in the
integrated SZ flux result.  For the two clusters above, A1835 and
A1689, increasing the integrating radius to $r_{500}$ also
increases the difference between $S(r_{500})$ derived from their
two respective published IC gas models.  For A1835 and A1689 the
magnitude of $S(r_{500})$ decreases by $\sim 6\%$ and $\sim16\%$,
respectively, when using the broader core radius Beta model,
relative to the narrower core radius model, for each cluster. This
difference is not surprising considering $r_{2500}$ is already
greater than the SuZIE beam-size for most of our clusters, see
Table \ref{tab:cluststuff}, with $r_{500}$ generally a factor of
$\sim$2 larger than $r_{2500}$ for a typical cluster. Regardless,
there seems to be a systematic trend towards over-estimating the
integrated SZ decrement out to $r_{500}$ by $\sim5$-$20\%$ when
using the narrower core radius models for cooling flow clusters.
This level of uncertainty is approximately equal to the
statistical uncertainty of our measurements, and should be
considered when extrapolating our integrated SZ flux measurements
to larger cluster radii.

\subsection{Systematic Effects from an Uncertain Electron Temperature}
\label{sec:teerror}

To examine the effect of an uncertain electron temperature on the
SuZIE results we consider two different possibilities.  First, we
consider a model with a non-isothermal temperature structure
implied from XMM measurements for one of our cooling flow
clusters, A1835.  Secondly, we consider the effect of a
significant systematic bias in the assumed isothermal electron
temperature.


\subsubsection{Non-Isothermal Temperature Structure}

A previous SuZIE paper by \citet{swlh97a} used a more complicated
cluster thermal structure, suggested from ASCA observations, to
analyze their 145 GHz results and found that the value of the
central Comptonization was relatively insensitive to the details
of the thermal structure.  With the current generation of X-ray
telescopes, XMM and Chandra, there exists a greater capability to
resolve cluster temperature structure. Of the clusters in our
sample, A1835 has several published X-ray results from XMM and
Chandra \citep[for example]{peterson, majer02}. In particular
\citet{majer02} used XMM observations to measure a radial
temperature profile in 6 annuli out to a radius of 6', or $\sim 3
r_{2500}$. Using a more realistic temperature profile for one of
our cooling flow clusters is of particular interest because of the
well-known drop in temperature in the central core of the cooling
flow.  To this end, we have adopted the radial temperature profile
given in \citet{majer02} and re-analyzed our results for A1835 to
study the effect of a non-isothermal temperature structure.

In Table \ref{tab:tera1835} we give the temperature profile of
A1835 given in \citet{majer02}.  The XMM observations did not have
sufficient sensitivity beyond a radius of 6' to significantly
measure the cluster temperature.  Little is known about the
temperature structure in the outer most regions of clusters,
although it is expected the temperature will decrease at some
radius. We will assume beyond 6' the cluster is isothermal with a
temperature of 7.6 keV, which is the mean temperature outside the
cooling flow calculated by \citet{majer02}.  While this is a
fairly arbitrary assumption, we will later show that the assumed
value only weakly affects our results.

In Table \ref{tab:a1835} we give the calculated central
Comptonization and the integrated SZ flux assuming the above
temperature profile and the Beta model parameters given in Table
\ref{tab:icmodel}.  We give results for both the multi-frequency
analysis and the 145 GHz analysis, which were described in
sections \ref{sec:ssfits} and \ref{sec:145y0} respectively.  For
comparison, we list the results from sections \ref{sec:ssfits} and
\ref{sec:145y0} which assumed an isothermal intra-cluster gas with
$kT_e=8.2$ keV. For both analysis methods the calculated central
Comptonization and integrated SZ flux change negligibly between
the above two temperature models compared to the statistical
uncertainty of our measurements. This suggests that our results
are largely insensitive to the detailed thermal structure of the
gas, and are particularly insensitive to the cool core in cooling
flow clusters. To examine the effect of the unknown temperature in
the outer region of the cluster, in Table \ref{tab:a1835} we also
give our results using the XMM thermal model but instead assuming
a temperature of 4.11 keV beyond a radius of 6 '. Comparing the
results assuming this model with those which assumed an isothermal
8.2 keV gas, the results again change negligibly compared to the
statistical uncertainty of our measurements.  In addition, the
results of the two XMM models in Table \ref{tab:a1835} are nearly
identical. Considering we changed the temperature of the outer
region by nearly a factor of two, this suggests that whatever
reasonable temperature we might assume for the outer region would
not have a significant effect on our results.  Therefore we
conclude that the thermal structure of the intra-cluster gas adds
a negligible uncertainty to our results.




\subsubsection{Systematic Bias in the Assumed Isothermal
Temperature}

We now consider the case of a significant systematic bias in the
assumed isothermal electron temperature greater than the quoted
statistical uncertainty. To some degree a systematic bias in the
electron temperature is expected due to any difference between the
X-ray derived temperature and the mass weighted temperature, which
is relevant for SZ observations. Simulations by \citet{ben01}
predict that temperatures derived from spectral fits to X-ray data
are $\sim1$-$3$keV less than the mass weighted temperature, with
the systematic offset proportional to the temperature of the
cluster. Here we discuss the effect of a systematic bias in the
temperature on the calculated central Comptonization and the
integrated SZ flux.


We can again consider the case of A1835, one of our more
significant detections, to examine the effect of a systematic
temperature uncertainty on the central Comptonization.
\citet{ben01} suggests that in the most extreme cases the mass
weighted temperature is $\sim$40\% higher than the X-ray spectral
temperature. For A1835 a 40\% offset in temperature corresponds to
a temperature of $\sim$11.5 keV.  In Table \ref{tab:a1835} we give
our results for A1835 using both the multi-frequency analysis and
the 145 GHz analysis, which were described in sections
\ref{sec:ssfits} and \ref{sec:145y0} respectively, assuming an
isothermal temperature of 11.5 keV.  Both the multi-frequency and
145 GHz results do not change significantly compared to the
results which assumed an isothermal temperature of 8.2 keV.  In
Table \ref{tab:a1835} we also give the results for A1835 assuming
an isothermal temperature of 4.11 keV, the temperature
\citet{majer02} measured for the cluster core.  While X-ray
observations strongly suggest that most of the cluster gas is well
above this temperature, our results do not significantly change
even with this most pessimistic assumption for the temperature.
Overall, in Table \ref{tab:a1835} our results change remarkably
insignificantly over a relatively extreme range of temperatures.
This insensitivity is because the calculated central
Comptonization depends on the temperature through relativistic
corrections to the SZ spectrum, which are still relatively small
at 145 GHz for a reasonable range of temperatures. We conclude
that any uncertainty in the electron temperature causes a
negligible contribution to the uncertainty of our results for all
the clusters in our sample.

We should note that if we had allowed $r_{2500}$ to vary with the
temperature according to equation \ref{eqn:r2500}, the results in
Table \ref{tab:a1835} for the integrated flux, $S(r_{2500})$,
would have changed significantly. Our definition of the
integration cut-off radius, $r_{2500}$, goes approximately like
$T_e^{1/2}$, which in turn causes a dependence of the integrated
SZ flux, $S(r_{2500})$, on temperature through the integral in
equation \ref{eqn:szint}.  Therefore a systematic bias in the
definition of the X-ray temperature across all our clusters would
tend to affect our integrated SZ flux results somewhat
proportionally. While this would have the potential to bias our
scaling relation calculations later in section \ref{sec:scaling},
it would not necessarily be evident in our relations through
excess scatter. Regardless, for comparison purposes it is useful
to use the same definition of $r_{2500}$ calculated from the X-ray
temperature which is used to calculate the analogous X-ray scaling
relations. For this reason we do not consider the effect of a
systematic uncertainty in the electron temperature on
$S(r_{2500})$ through the definition of $r_{2500}$, except to note
that a systematic bias could exist in the X-ray determined
electron temperatures, and generally should be considered when
interpreting our results.

\section{Comparison of SZ Measurements with Previous
Results} \label{sec:y0compare}

In total the SuZIE observing program has detected the SZ spectrum
of 13 clusters of galaxies and detected an SZ decrement in 15
clusters of galaxies, see \citet{swlh97b,benson}, and this paper.
An important systematic check is to compare our results to SZ
measurements using other instruments.  The most comprehensive set
of SZ measurements published are those by \citet{reese02} using
the Berkeley-Illinois-Maryland Association (BIMA) and Owens-Valley
Radio Observatory (OVRO) millimeter wavelength interferometers. A
total of 10 clusters overlap between the SuZIE cluster sample and
the set published in \citet{reese02}. In this section we derive a
central Comptonization for each of the 10 overlapping clusters
published in \citet{reese02} and compare these values for the
calculated central Comptonizations from SuZIE.  To simplify the
comparison of these clusters we have used the same IC gas model as
\citet{reese02} to analyze our measurements.

\subsection{BIMA and OVRO} \label{sec:bimaovro}

The BIMA and OVRO arrays are millimeter wavelength
interferometers, which have been outfitted with centimeter
wavelength receivers to observe the SZ effect.  The receivers use
High Electron Mobility Transistor (HEMT) amplifiers which are used
to observe the SZ effect in a band between 28-30 GHz. At this
observing frequency the primary beams for each interferometer are
nearly Gaussian with a FWHM of 6.6' for BIMA and 4.2' for OVRO.
The angular resolution varies depending on the configuration of
the dishes during each particular observation, but is typically
$\sim95\times95$ arcsec for BIMA and $\sim50\times50$ arcsec for
OVRO.  For an overview of the interferometers and the SZ
observations using them see \citet{reese02}.

\subsubsection{Fitting a Central Comptonization to the BIMA and
OVRO Data} \label{sec:bimaovroy0}

The BIMA and OVRO interferometers observe the SZ effect in a
narrow frequency band at $\nu \sim 28.5$ GHz.  Because their
measurements are effectively at a single frequency they are unable
to constrain both a central Comptonization and a peculiar velocity
from their data alone.  Instead, the SZ results quoted by
\citet{reese02} give the central intensity of each cluster in
units of thermodynamic temperature.  The measured difference
temperature, $\Delta T$, is then a sum of thermal and kinematic
components with
\be \Delta T = y_0 T_{\rm CMB} \frac{(e^x-1)^2}{x^4 e^x} \frac{m_e
c^2}{kT_e} \left[ \Psi(x,T_e)-\frac{v_p}{c} h(x,T_e) \right]
\label{eqn:ovrodeltaT} \ee
where $\Psi(x,T_e)$ and $h(x,T_e)$ are fully specified in
\citet{benson} and include relativistic corrections to their
frequency dependence based on the calculations of \citet{yoel} and
\citet{relkin} respectively.

In order to compare the BIMA and OVRO measurements to ours we want
to fit a central Comptonization to the temperature decrement given
in \citet{reese02}. The observations described in \citet{reese02}
were taken in two different receiver configurations with central
observing frequencies of $\nu=28.5$ and $30.0$ GHz respectively.
\citet{reese02} does not systematically note which sets of data
correspond to which observing frequency. Because we are nearly in
the Rayleigh-Jeans region of the spectrum, the calculated central
Comptonization for a typical cluster varies by $< 1\%$ if we
assume a central observing frequency between 28.5 and 30.0 GHz.
For simplicity, we therefore assume all observations in
\citet{reese02} were taken at a central observing frequency of
$\nu=28.5$ GHz with a Gaussian envelope 0.5 GHz in width. We can
then calculate a central Comptonization derived from the published
central decrements in \citet{reese02} in a way exactly analogous
to the method in section \ref{sec:145y0}, which was used to
analyze the SuZIE 145 GHz data.  From equation
\ref{eqn:ovrodeltaT}, we calculate a two-dimensional
$\chi^2(v_p,y_0)$ over an appropriate range of parameter space for
peculiar velocity and central Comptonization. Under the assumption
of Gaussian errors on $\Delta T$, we calculate a likelihood,
$L(v_p,y_0) \propto \exp(-\chi^2(v_p,y_0)/2)$. We multiply
$L(v_p,y_0)$ by a Gaussian prior on the peculiar velocity, where
$L(v_p) \propto \exp(-v_p^2/2\sigma_v^2)$, with $\sigma_v=0,500,$
and $2000$ km s$^{-1}$ as our three cases.  We then marginalize
the resultant likelihood over peculiar velocity to calculate the
best-fit Comptonization and 68\% confidence region for the three
cases of $\sigma_v$ and give these results in Table
\ref{tab:y0fits}

In general the main effect of increasing the width of the Gaussian
prior on the peculiar velocity is to expand the corresponding
confidence region for the central Comptonization.  For each
cluster the BIMA and OVRO best-fit central Comptonization changes
negligibly between the different priors, however the width of the
confidence region expands by a factor of $\approx 2-3$ between an
exactly zero peculiar velocity and $\sigma_v=2000$\,km\,s$^{-1}$.


\subsection{Comparing SuZIE to BIMA/OVRO}
\label{sec:suzieovrocompare}

We can compare the central Comptonization results calculated from
BIMA/OVRO to the central Comptonization calculated from both the
SuZIE multi-frequency data and the SuZIE 145 GHZ data.  The SuZIE
145 GHz data gives better constraints on the central
Comptonization than the multi-frequency results, for reasons
discussed in section \ref{sec:145y0}, however both comparisons are
useful because they are sensitive to different systematics. For
example, the SuZIE multi-frequency results may be more appropriate
if clusters have larger peculiar velocities than expected,
conversely the SuZIE 145 GHz data would be more appropriate if
sub-millimeter point sources bias the higher frequency channels.

To facilitate comparison to the BIMA/OVRO results, we re-analyze
the multi-frequency SuZIE results from section \ref{sec:ssfits}
using the same method used to analyze the BIMA/OVRO results as
described in the previous section. In Table \ref{tab:y0fits} we
give the best-fit Comptonization and 68\% confidence region
derived from the multi-frequency SuZIE results assuming three
different priors on the peculiar velocity with Gaussian widths of
$\sigma_v=0,500,$ and $2000$ km s$^{-1}$. We note that we are
considering a broader range of priors on the peculiar velocity on
the SuZIE multi-frequency results versus the SuZIE 145 GHz
results.  We do this because the multi-frequency results already
constrain the peculiar velocity to some degree and therefore
should be more appropriate if we are considering a larger range of
possible peculiar velocities.

Comparing the results of Table \ref{tab:y0fits}, the SuZIE derived
central Comptonizations are higher than the results from BIMA and
OVRO for all the clusters except Cl0016. For the case of $\sigma_v
= 500$ km s$^{-1}$, the clusters A697, A773, RXJ147, A1835, and
A2261 all have significantly higher Comptonizations as measured by
SuZIE. Even for the case of $\sigma_v = 2000$ km s$^{-1}$, the
clusters A1835 and A2261 are still significantly inconsistent
between the two data sets.

\begin{figure}
\plotone{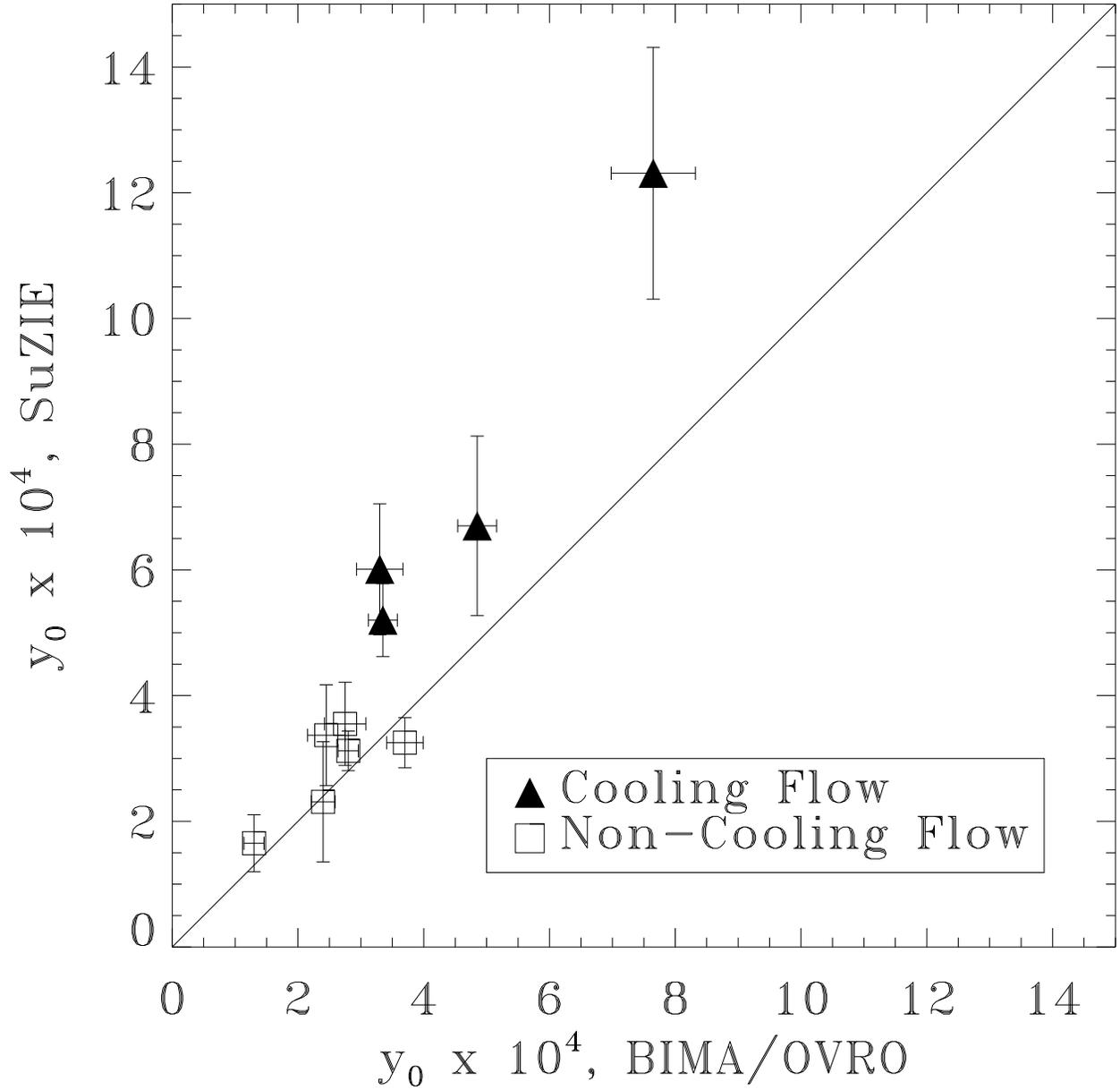} \caption[]{A plot of the central Comptonization
measured by SuZIE versus the central Comptonization measured by
the BIMA/OVRO interferometers.  The SuZIE central Comptonization
calculation is based on the method described in section
\ref{sec:145y0}.  Both measurements assume a zero peculiar
velocity with a Gaussian prior on the peculiar velocity with a
width of 500 km s$^-1$. Clusters with cooling cores are labelled
with triangles and non-cooling core clusters are labelled with
squares.} \label{fig:y0vy0}
\end{figure}

If instead we compare the central Comptonizations calculated from
the SuZIE 145 GHz data in section \ref{sec:145y0} to the BIMA and
OVRO results, SuZIE continues to measure a higher central
Comptonization for most clusters.  Figure \ref{fig:y0vy0} plots
the central Comptonization calculated from the SuZIE 145 GHz data,
calculated in section \ref{sec:145y0}, to the central
Comptonization calculated from the BIMA and OVRO measurements,
where we have assumed $\sigma_v = 500$ km s$^{-1}$ for both
calculations. Again the SuZIE derived central Comptonizations are
systematically higher than the OVRO/BIMA results, particularly in
cooling flow clusters. On average, the SuZIE calculated central
Comptonizations are $\sim12\%$ higher in the non-cooling flow
clusters, and $\sim60\%$ higher in the cooling flow clusters.

This discrepancy is equivalent to the statement that SuZIE is
measuring a systematically higher SZ flux than expected for the
spherical isothermal beta model normalized to the central
Comptonization given in \citet{reese02}.  This suggests that the
SuZIE measurement is either inconsistent with this central
Comptonization, and/or the IC gas model.  It was shown in section
\ref{sec:145y0}, that the central Comptonization derived by SuZIE
is very sensitive to the assumed IC gas model.  Given that SuZIE
is sensitive to different spatial scales than OVRO or BIMA, it is
possible that the SuZIE measurements be consistent with the OVRO
and BIMA measurements and still derive a different central
Comptonization if the Beta model does not fit the IC gas
distribution well. This discrepancy will be investigated further
in a future paper.

\section{SZ Scaling Relations}
\label{sec:scaling}

Self-similar models of cluster formation, which include only
gravity and shock heating, predict scaling relations between the
electron temperature, integrated SZ flux, and central
Comptonization, \citep[see][for example]{kaiser86, navarro95,
dasilva03}. In the self-similar model the mass and temperature of
a cluster are related by $M E(z) \propto T^{3/2}$, where
$E(z)^2\equiv \Omega_M (1+z)^3 +
(1-\Omega_M-\Omega_{\Lambda})(1+z)^2 + \Omega_{\Lambda}$.  The
factor of $E(z)$ arises from the assumption that the cluster
density scales with the mean density of the Universe. Following
\citet{dasilva03}, the mass-temperature scaling relation can be
used to relate the SZ flux, $S$, to the temperature
\be S d_A(z)^2 E(z) \propto T^{5/2} \label{eqn:svt_ss} \ee
where $d_A(z)$ is the angular diameter distance to the cluster.
The factor of $d_A(z)^2$ accounts for the apparent angular size of
the cluster, which changes with the cluster's redshift.  The
scaling of the central Comptonization with the temperature can be
derived through its relation to the integrated SZ flux
\be S = \int \Delta I d\Omega \propto y_0 \int d\Omega \propto
\frac{y_0}{d_A^2} \int dA \ee
where $dA$ corresponds to a physical radius such that $\int dA =
\pi r^2$, where $r$ is the radius of the cluster. The radius of
the cluster can be related to the cluster mass, $M$, and the
critical density of the universe, $\rho_{\rm crit}$, by equation
\ref{eqn:clustmass}, such that $r^3 \propto M / \rho_{\rm crit}
\propto T^{3/2} / E(z)^3$, where we have used the mass-temperature
relation, $M E(z) \propto T^{3/2}$. Combining this result with the
SZ flux-temperature scaling relation, and the definition of the
central Comptonization above, we arrive at
\be \frac{y_0}{E(z)} \propto T^{3/2}  \label{eqn:yvt_ss} \ee
as the expected scaling between the central Comptonization and the
temperature of the cluster.  
The published temperatures that we use are spectral temperatures
derived from fits to X-ray spectra.  As previously mentioned,
simulations by \citet{ben01} predict that temperatures derived
from spectral fits to X-ray data would be $\sim1$-$3$keV less than
the mass weighted temperature. \citet{ben01} calculated the effect
of using the spectral temperature in place of the mass-weighted
temperature in the mass-temperature scaling relation and found $M
E(z) \propto T^{1.6}$.  Because we are using spectral temperatures
when calculating our scaling relations, we expect equations
\ref{eqn:svt_ss} and \ref{eqn:yvt_ss} to be proportionally
steepened.
%
%
%

The predicted slopes and offsets of the above scaling relations
also change from the presence of other cooling and heating
processes. Heat input, through sources such as radiative cooling
or pre-heating, steepens the slope of the mass-temperature,X-ray
luminosity-temperature, and SZ flux-temperature scaling relations
\citep[see][for example]{verde02, dasilva03}. The steepening of
the X-ray luminosity-temperature and mass-temperature relations
have been observed by several authors using X-ray measurements
\citep[see][for example]{mark98,fin01}.  In particular,
\citet{fin01} found that $M \propto T_X^{1.78^{+0.10}_{-0.09}}$
(68\%) from X-ray observations of relatively nearby ($z \lesssim
0.1$) clusters, which is significantly steeper than the
self-similar predicted slope of 1.5. Little work has been done to
measure SZ scaling relations due to the scarcity of SZ
measurements. \citet{cooray99} and \citet{mccarthyb} have compiled
SZ measurements from the literature, but concentrated almost
entirely on relations using the central decrement, which as we
have shown could be susceptible to significant systematic
uncertainties. No measurements exist of an integrated SZ
flux-temperature relation. This study is also the first which use
results entirely from one instrument to construct SZ scaling
relations.

\subsection{Definition of the Fit} \label{sec:logfit}

To fit the following relations we perform a linear least squares
regression in log space.  Uncertainties for an arbitrary variable
$X$ are transformed into log space by the relation $\sigma_{log
(X)} = (X^+-X^-)/(2X) \times {\rm log}(e)$ where $X^+$ and $X^-$
are the positive and negative errors, respectively, to the
variable $X$. We perform a linear least squares regression to the
generic relation ${\rm log}(Y) = A + B {\rm log}(X)$ where we
determine the best-fit values of $A$ and $B$ by minimizing our
$\chi^2$ statistic which we define as
\be  \chi^2 = \sum_{i=1}^{N} \frac{\left({\rm log}(Y_i) - B {\rm
log}(X_i) - A\right)^2} {\sigma_{{\rm log}(Y_i)}^2+(B \sigma_{{\rm
log}(X_i)})^2} \label{eqn:logchi} \ee 
%
where $\sigma_{{\rm log}(X_i)}$ and $\sigma_{{\rm log}(Y_i)}$ are
the uncertainties to $X_i$ and $Y_i$, respectively, transformed
into log space, as defined above, for the $i$th cluster.  The
uncertainties on $A$ and $B$, $\sigma_A$ and $\sigma_B$, are
defined in a standard way using a general definition from a linear
least squares fit \citep[see][for example]{press}.

\subsection{$S(r_{2500}) d_A^2 E(z)$--$T_X$}

\begin{figure}
\plotone{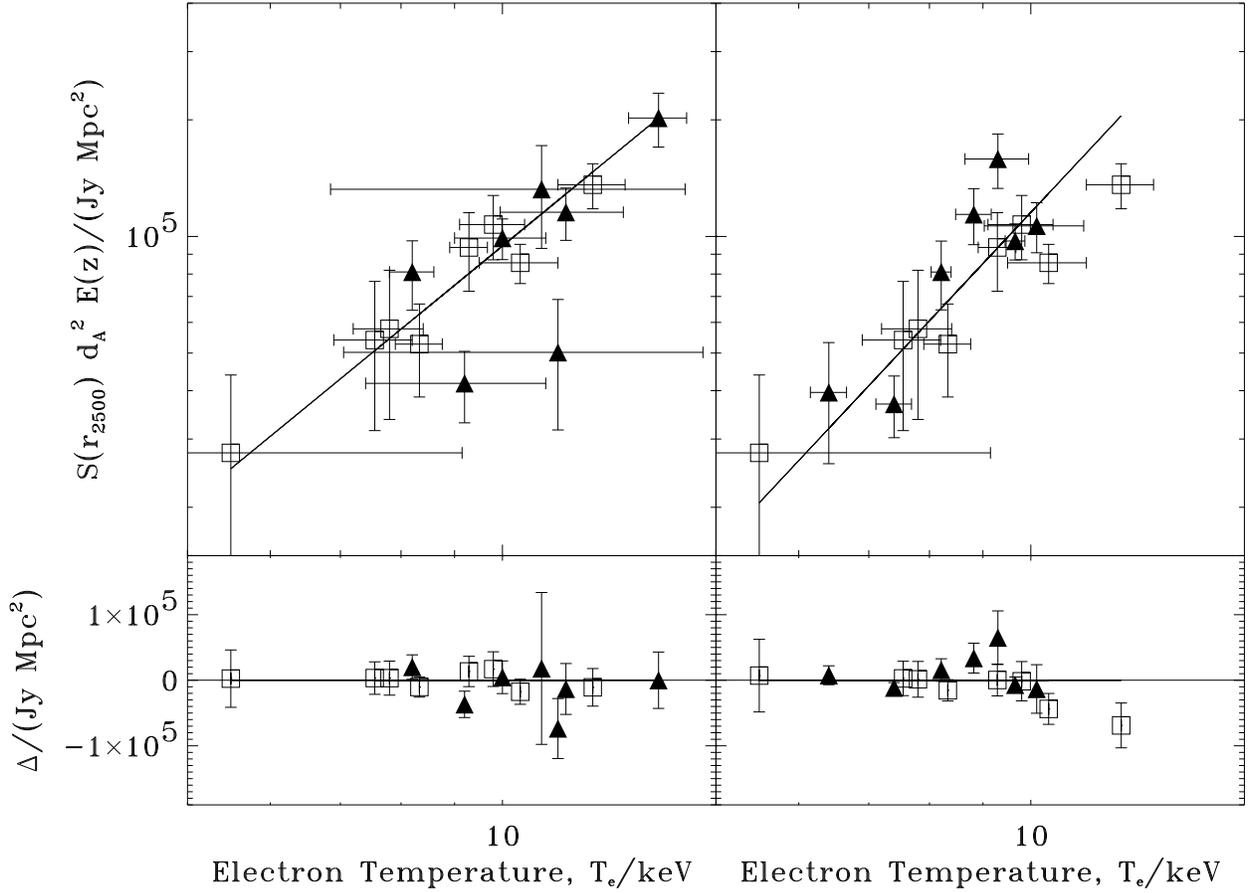} \caption[]{Left:(Top) A plot of the integrated SZ
flux, as measured by SuZIE, versus the electron temperature. The
solid line shows a power-law fit to the relation. Cooling flow
clusters are plotted as triangles, and non-cooling flow clusters
are plotted as squares.  (Bottom) A plot of the residuals to the
power-law fit. The uncertainty on electron temperature is not
plotted, but instead is added in quadrature, according to equation
\ref{eqn:svte}, with the uncertainty to the flux density to give
the uncertainty for the residual data points. Right: The same plot
as on the left, except for the cooling flow clusters we use
electron temperatures which do not account for the presence of the
cooling flow in our calculation of $S(r_{2500})$.}
\label{fig:szintvte}
\end{figure}

In this section we construct an integrated SZ flux, $S(r_{2500})$,
versus X-ray temperature, $T_X$, scaling relation.  The values we
use for $S(r_{2500})$, $E(z)$, $T_X$, and $d_A$ are given in Table
\ref{tab:cluststuff}. Where relevant, the data in Table
\ref{tab:cluststuff} assumes the cooling flow corrected
temperature. According to the method described in section
\ref{sec:logfit}, we fit a line to ${\rm log}[S(r_{2500}) d_A(z)^2
E(z)]$ versus ${\rm log}[T_X]$ whose best-fit relationship is
\be {\rm log}\left[\frac{S(r_{2500}) d_A^2 E(z)}{\rm Jy
Mpc^2}\right] = (2.76 \pm 0.41) + (2.21\pm0.41) {\rm
log}\left[\frac{T_X}{\rm keV}\right] \label{eqn:svte} \ee
where the error bars correspond to the 68\% confidence region for
both the offset and slope.  The $\chi^2$ to the fit is 6.52 for 13
degrees of freedom.  This low $\chi^2$ implies that we are not
seeing any sources of intrinsic scatter in the relation, and are
currently limited by measurement uncertainty. Figure
\ref{fig:szintvte} plots $S(r_{2500}) d_A(z)^2 E(z)$ versus $T_X$
for the entire 15 cluster sample with the best-fit line from
equation \ref{eqn:svte} over-plotted.  The best-fit slope is
slightly less than the expected self similar slope of 2.5, see
equation \ref{eqn:svt_ss}, however it is well within the 68\%
confidence region.  X-ray measurements suggest a steeper
mass-temperature relation which would also imply a steeper slope
approximately between 2.7-2.9 for the integrated SZ
flux-temperature relation. Our results suggest a smaller slope,
however they lack the sensitivity to say anything significant
regarding this difference.

It is also of interest to consider any systematic difference
between the cooling flow and non-cooling flow sub-samples. In
Table \ref{tab:szvtefits} we show the results of the fits to
equation \ref{eqn:svte} if we consider the cooling flow and
non-cooling flow sub-samples separately.  The best-fit lines for
the two sub-samples are nearly identical, and almost unchanged to
the best-fit line for the entire sample.  This suggests either
that the presence of cooling flows make a negligible correction to
the SZ flux-temperature scaling relation, or that the temperature
we are using have accurately corrected for the presence of the
cooling flows. We can test which is the case by re-calculating the
scaling relation using the cooling-flow uncorrected temperatures.
We do this by re-calculating $S(r_{2500})$, as prescribed in
section \ref{sec:szint}, instead assuming the X-ray emission
weighted temperatures in Table \ref{tab:icmodel}, which do not
account for the cooling flow. The right panel of Figure
\ref{fig:szintvte} re-plots $S(r_{2500}) d_A(z)^2 E(z)$ versus
$T_X$ using the re-calculated values of $S(r_{2500})$ with these
different temperatures. Comparing the left to the right panel of
Figure \ref{fig:szintvte}, only the points for the cooling flow
clusters are changed, with the cooling flow clusters in the right
panel having generally lower electron temperatures because they do
not account for their cool cooling core in their determination of
the electron temperature.  In Table \ref{tab:szvtefits}, we give
the new best-fit lines for the entire 15 cluster sample, and then
the cooling flow and non-cooling flow sub-samples separately.  The
best-fit line which describes the cooling flow clusters is
significantly changed between the cooling flow un-corrected and
cooling flow corrected temperatures.  This suggests that the
presence of the cooling flow needs to accounted for in calculating
the electron temperature in order to accurately measure the
$S(r_{2500}) d_A^2 E(z)$--$T_X$ scaling relation.


\subsubsection{Measuring the Evolution of the $S(r_{2500}) d_A^2 E(z)$--$T_X$
Relation}

The temperature of the intra-cluster gas is expected to scale with
$E(z)$ based on the assumption that the density of the gas scales
with the mean density of the universe.  As was mentioned at the
beginning of this section, this effect causes a redshift evolution
in the mass-temperature relation such that $M E(z) \propto
T^{3/2}$.  The redshift evolution of this relation was constrained
by X-ray observations in \citet{vikhlinin}. A similar redshift
evolution is expected in the SZ flux-temperature relation, such
that $S d_A(z)^2 E(z) \propto T^{5/2}$, as a direct consequence of
the redshift evolution in the mass-temperature relation.  However,
other non-gravitational physics could affect this predicted
redshift evolution.

Recently \citet{dasilva03} used numerical simulations to study the
evolution of the integrated SZ flux versus X-ray temperature
relation when including other non-gravitational effects in
clusters, such as from radiative cooling or pre-heating of the
intra-cluster gas. They parameterized an arbitrary evolution by
assuming that the temperature of the intra-cluster gas scaled like
$E(z)^{\gamma}$ and then fit for $\gamma$ using simulated clusters
which included either radiative cooling or pre-heating. They
calculated $\gamma=1.49$ in their simulations which included
radiative cooling and $\gamma=1.22$ in their simulations which
included pre-heating instead. To fit our data we adopt a similar
approach to \citet{dasilva03} and fit the relation
\be {\rm log}\left[\frac{S(r_{2500}) d_A^2 E(z)^{\gamma}}{\rm Jy
Mpc^2}\right] = A + B {\rm log}\left[\frac{T_X}{\rm keV}\right]
\label{eqn:szintevol} \ee
while allowing $\gamma$ to be a free-parameter, where we have
assumed the cooling-flow corrected electron temperature, for the
cooling flow clusters, in our calculation of $S(r_{2500})$ and
$T_X$.  We calculate the $\chi^2$ of the fit to equation
\ref{eqn:szintevol} for a range of $\gamma$, letting the offset
and slope, $A$ and $B$, go to their best-fit values for each value
of $\gamma$.  We then calculate our best-fit value of $\gamma$ and
its associated confidence regions using a maximum likelihood
estimator, where $L(\gamma) \propto \exp(-\chi^2(\gamma)/2)$.
Doing this we calculate $\gamma = 1.16^{+0.84 +1.28}_{-0.71
-1.14}$, where the uncertainties correspond to the 68\% confidence
region followed by the 90\% confidence region.  Our results do not
have sufficient sensitivity to significantly favor either of the
models of \citet{dasilva03}.  However, we can rule out at $\sim
90\%$ confidence zero evolution to the integrated SZ
flux-temperature relation; this is the first constraint of any
kind on the redshift evolution of this relation.  Furthermore, the
redshift evolution we observe is consistent with standard theories
of cluster formation ($\gamma=1$), and offers indirect evidence
regarding the redshift evolution of the mass-temperature relation.

\subsection{$y_0/E(z)$--$T_X$}

In this section we construct a central Comptonization, $y_0$,
versus X-ray temperature, $T_X$, scaling relation.  We showed in
section \ref{sec:systic} that the central Comptonization had a
significant systematic uncertainty due to the modelling of the IC
gas distribution.  Therefore we would expect this systematic
uncertainty to make any scaling relation involving the central
Comptonization suspect at best.  However, it may be interesting to
see how this systematic uncertainty manifests itself in a
$y_0/E(z)$--$T_X$ scaling relation.

\begin{figure}\epsscale{0.75}
\plotone{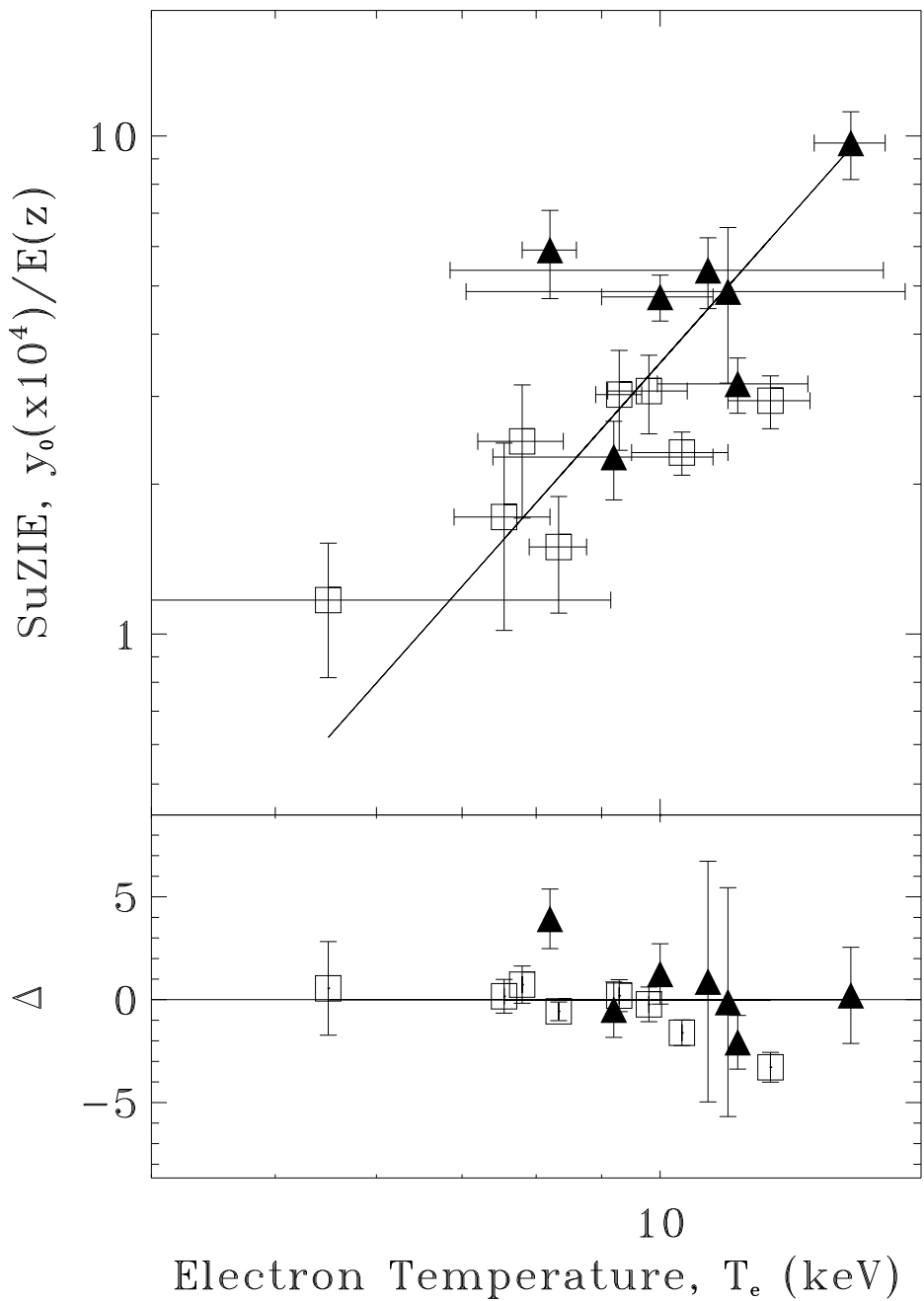} \caption[]{(Top) A plot of the central
Comptonization, as measured by SuZIE, versus the electron
temperature. The solid line shows a power-law fit to the relation.
Cooling flow clusters are plotted as triangles, and non-cooling
flow clusters are plotted as squares. (Bottom) A plot of the
residuals to the power-law fit. The uncertainty on electron
temperature is not plotted, but instead is added in quadrature
with the uncertainty to the flux density to give the uncertainty
for the residual data points.} \label{fig:y0tx}
\end{figure}

To construct a $y_0/E(z)$--$T_X$ scaling relation we use the
central Comptonizations calculated in section \ref{sec:145y0} and
the X-ray temperatures, $T_X$, given in Table
\ref{tab:cluststuff}. Where relevant, the data in Table
\ref{tab:cluststuff} assumes the cooling flow corrected
temperature. We fit a line to ${\rm log}(y_0/E(z))$ versus ${\rm
log}(T_X/{\rm keV})$, according to the method described in section
\ref{sec:logfit}, whose best-fit relationship is
\be {\rm log}\left[\frac{y_0}{E(z)}\right] = (-2.35 \pm 0.57) +
(2.90\pm0.57) {\rm log}\left[\frac{T_X}{\rm keV}\right]
\label{eqn:yvte} \ee
where the error bars correspond to the 68\% confidence region for
both the offset and slope.  The $\chi^2$ to the fit is 38.0 for 13
degrees of freedom, with the $\chi^2$ dominated by the
contribution from A1835. Figure \ref{fig:y0tx} plots $y_0/E(z)$
versus $T_X$ for the entire 15 cluster sample with the best-fit
line from equation \ref{eqn:yvte} over-plotted.  To check the
effect of A1835 on the overall fit, we refit equation
\ref{eqn:yvte} excluding A1835, with these results given in Table
\ref{tab:yvtefits}. Excluding A1835 negligibly changes the
best-fit values for the slope and offset while reducing the
$\chi^2$ to 15.0 for 12 degrees of freedom. This seems to indicate
our fit of the $y_0$--$T_X$ scaling relation is reasonable,
however the best-fit slope in equation \ref{eqn:yvte} is
inconsistent with the self-similar prediction of 1.5.

If we consider the cooling flow and non-cooling flow clusters
separately there is a significant systematic difference between
them. In Table \ref{tab:yvtefits} we show the results of the fits
to equation \ref{eqn:yvte} if we consider the cooling flow and
non-cooling flow sub-samples separately.  The best-fit line to the
cooling flow clusters actually favors a negative slope but is
clearly poorly constrained.  If we exclude A1835 from the cooling
flow sample, and refit the remaining clusters, we calculate a
best-fit line which is consistent with the non-cooling flow
sub-sample, however the constraints on the slope are very poor.
The best-fit line to the non-cooling flow sub-sample has a slope
marginally consistent with the self-similar prediction. From
Figure \ref{fig:y0tx} it is clear that the cooling flow sub-sample
is not well fit by a line. This is not surprising considering that
in section \ref{sec:145y0} we showed that the central
Comptonization has a large systematic dependence on the assumed
spatial gas distribution for cooling flow clusters. The
non-cooling flow sub-sample visibly gives a better fit to a line
than the cooling flow sub-sample, however it is difficult to
ascertain the degree of systematic uncertainty in this relation.



\section{Conclusions}

We report new measurements of the SZ effect from clusters of
galaxies in three frequency bands. We use the multi-frequency
measurements to measure the SZ spectrum and the central
Comptonization in each cluster.  We combine these new measurements
with previous cluster observations to construct a sample of 15
clusters of galaxies detected with the SuZIE experiment. For the
entire set of clusters we use the 145 GHz frequency band to
calculate a central Comptonization, $y_0$, and an integrated SZ
flux, $S(r_{2500})$.

We find that the calculated central Comptonization is much more
sensitive to the assumed spatial model for the intra-cluster gas
than the calculated integrated SZ flux.  The calculated central
decrement depends significantly on the assumed spatial
distribution of the intra-cluster (IC) gas. For the case of A1835
the calculated central Comptonization can vary by a factor of two
depending on which of two different published IC gas models is
assumed.  This result is not surprising considering the fact that
SuZIE~II does not significantly resolve any of the observed
clusters. This effect causes the calculated central Comptonization
to have a particularly large systematic uncertainty in cooling
flow clusters because of their large cooling core which makes the
standard Beta model an inadequate fit to the spatial distribution
of the gas. However, our measurements of the integrated SZ flux,
$S(r_{2500})$, negligibly depend on the assumed spatial
distribution of the IC gas because $r_{2500}$ is well-matched to
our beam-size for most of the observed clusters.


Ten of the clusters in our sample overlap with published
measurements from the BIMA and OVRO interferometers.  For these
clusters, we compare the calculated central Comptonization from
BIMA and OVRO to those from SuZIE and find that the SuZIE
calculated central Comptonizations are generally higher,
significantly so in the cooling flow clusters.  If we compare the
SuZIE 145 GHz results to the BIMA and OVRO results, SuZIE measures
a central Comptonization $\sim12\%$ higher in the non-cooling flow
clusters, and $\sim60\%$ higher in the cooling flow clusters.  We
attribute this difference to the large systematic uncertainty in
the calculated central Comptonization from the assumed
intra-cluster gas model which, as expected, is more pronounced in
our cooling flow sub-sample.

We use the central Comptonization and integrated SZ flux results
from the SuZIE 145 GHz data to construct SZ scaling relations with
the X-ray temperature, $T_X$. We construct a $y_0$--$T_X$ scaling
relation and find a slope significantly different than what is
expected for self-similar clusters.  However, we believe that this
result is questionable because of the large systematic uncertainty
in the central Comptonization.  This conclusion is supported by
the significantly discrepant scaling relations derived for the
cooling flow and non-cooling flow sub-samples. For the
$S(r_{2500})$--$T_X$ scaling relation we find a slope which is
consistent with the expectation for self-similar clusters. In
constructing this relation, we find that using X-ray temperatures
which do not account for the presence of the cooling flow
significantly biases the best-fit relation. We detect a redshift
evolution of the $S(r_{2500})$--$T_X$ scaling relation consistent
with standard cluster formation theory for which the density of
the cluster scales with the mean density of the universe.  If we
assume that the X-ray temperature is a good indicator of the mass
of the cluster, as suggested from X-ray measurements, our results
imply that the the integrated SZ flux will be a good indicator of
the cluster mass as well, a promising result for future SZ cluster
surveys.

We thank Jim Bartlett, Gilbert Holder, and Phil Marshall for many
useful discussions.  The SuZIE program is supported by a National
Science Foundation grant AST-9970797, NASA grant NAG5-12973, and a
Stanford Terman Fellowship awarded to SEC. This work was partially
carried out at the Infrared Processing and Analysis Center and the
Jet Propulsion Laboratory of the California Institute of
Technology, under a contract with the National Aeronautics and
Space Administration. We would like to thank the CSO staff for
their assistance with SuZIE observations. The CSO is operated by
the Caltech Submillimeter Observatory under contract from the
National Science Foundation.

\newpage

\begin{deluxetable}{lcccccccc}
\tablecaption{Summary of New SuZIE observations\label{tab:obs}}
\tablewidth{0pt}
  \tablehead{
& & \colhead{R.A.\tablenotemark{a}} & \colhead{Decl.\tablenotemark{a}} & & \colhead{Total} & \colhead{Accepted} & \colhead{Integration} \\
\colhead{Source} & \colhead{z} & \colhead{(J2000)} &
\colhead{(J2000)} & \colhead{Date} & \colhead{Scans} &
\colhead{Scans} & \colhead{Time (hours)} & \colhead{Ref.}
    }
 \startdata
    A520  & 0.199 & 04 54 07.4 & $+$02 55 12.1 & Dec 02 & 462 & 435 & 14.5 & 1\\
    A545 & 0.153 & 05 32 23.3 & $-$11 32 09.6 & Dec 98 & 295 &266 & 8.9 & 3\\
    A697 & 0.282 & 08 42 57.8 & $+$36 21 54.0 & Mar 03 & 258 & 254 & 8.5 & 1\\
    A773 & 0.217 & 09 17 52.1 & $+$51 43 48.0 & Mar 03 & 92 & 83 & 2.8 & 1 \\
    MS1054 & 0.823 & 10 56 58.6 & $-$03 37 36.0 & Jan 02 & 236 & 219 & 7.0 & 4\\
    RXJ1347 & 0.451 & 13 47 31.0 & $-$11 45 11.0 & Mar 03 & 209 & 202 & 6.7 & 2\\
    A2204 & 0.152 & 16 32 47.0 & $+$05 34 33.0 & Mar 03 & 468 & 449 & 15.0 & 1\\
    A2390 & 0.232 & 21 53 36.7 & $+$17 41 43.7 & Dec 02 & 203 & 195 & 6.5 & 1\\
\enddata
 \tablerefs{(1) \citet{1998MNRAS.301..881E}(2) \citet{schindler}(3) \citet{1996MNRAS.281..799E}(4) \citet{1994ApJS...94..583G}}
 \tablenotetext{a}{Units of RA are
hours, minutes and seconds and units of declination are degrees,
arcminutes and arcseconds}
\end{deluxetable}

\begin{deluxetable}{lcccccl}
\tablecaption{IC gas temperatures and $\beta$ model parameters
\label{tab:icmodel}}\tablewidth{0pt} \tablehead{
  & \colhead{$kT_e$\tablenotemark{a}} & \colhead{$kT_e$\tablenotemark{b}} &  & \colhead{$\theta_{c}$} &  & \\
  \colhead{Cluster} & \colhead{(keV)} & \colhead{(keV)} & \colhead{$\beta$} &
  \colhead{(arcsec)} & CF or NCF & \colhead{Ref.} }
 \startdata
  A520 &  $8.33^{+0.46}_{-0.40}$ & \nodata  & $0.844^{+0.040}_{-0.040}$ & $123.3^{+8.0}_{-8.0}$ &  NCF & 1;2;2\\
  A545 &  $5.50^{+6.2}_{-1.1}$ & \nodata & $0.82$\tablenotemark{c} & $115.5$\tablenotemark{c} & NCF& 3;4;4\\
  A697 &  $9.8^{+0.7}_{-0.7}$ & \nodata & $0.540^{+0.045}_{-0.035}$ & $37.8^{+5.6}_{-4.0}$ & NCF& 2;2;2\\
  A773 &  $9.29^{+0.41}_{-0.36}$ & \nodata & $0.597^{+0.064}_{-0.032}$ & $45.0^{+7.0}_{-5.0}$ & NCF& 1;2;2\\
  MS1054 &  $7.8^{+0.6}_{-0.6}$ & \nodata & $1.39^{+0.14}_{-0.14}$ & $67.7$\tablenotemark{c} & NCF& 5;5;5\\
  RXJ1347 & $9.3^{+0.7}_{-0.6}$  & $14.1^{+0.9}_{-0.9} $ & $0.604^{+0.011}_{-0.012}$ & $9.0^{+0.5}_{-0.5}$ &  CF& 6;5;2;2 \\
  A2204 & $7.4^{+0.30}_{-0.28}$ & $9.2^{+2.5}_{-1.1} $ & $0.66$\tablenotemark{c} & $34.7$\tablenotemark{c} & CF& 1;1;4;4 \\
  A2390 &  $10.13^{+1.22}_{-0.99}$ & $11.5^{+1.5}_{-1.6}$ & $0.67$\tablenotemark{c} & $52.0$\tablenotemark{c} & CF& 1;7;4;4 \\
  A2261 &  $8.82^{+0.37}_{-0.32}$ & $10.9^{+5.9}_{-2.2} $ & $0.516^{+0.014}_{-0.013}$ & $15.7^{+1.2}_{-1.1}$ & CF& 1;1;2;2\\
  Zw3146 &  $6.41^{+0.26}_{-0.25}$ & $11.3^{+5.8}_{-2.7} $ & $0.74$\tablenotemark{c} & $13.0$\tablenotemark{c} & CF& 1;1;4;4 \\
  A1835 & $8.21^{0.19}_{-0.17}$ & $8.2^{+0.4}_{-0.4} $ & $0.595^{+0.007}_{-0.005}$ & $12.2^{+0.6}_{-0.5}$ & CF& 1;8;2;2 \\
  Cl0016 & $7.55^{+0.72}_{-0.58}$ & \nodata & $0.749^{+0.024}_{-0.018}$ & $42.3^{+2.4}_{-2.0}$ & NCF& 9;2;2 \\
  MS0451 &  $10.4^{+1.0}_{-0.8}$ & \nodata & $0.806^{+0.052}_{-0.043}$ & $34.7^{+3.9}_{-3.5}$ & NCF& 10;2;2 \\
  A1689 &  $9.66^{+0.22}_{-0.20}$ & $10.0^{+1.2}_{-0.8} $ & $0.609^{+0.005}_{-0.005}$ & $26.6^{+0.7}_{-0.7}$ & CF& 1;1;2;2 \\
  A2163 &  $12.2^{+1.1}_{-0.7}$ & \nodata & $0.674^{+0.011}_{-0.008}$ & $87.5^{+2.5}_{-2.0}$ & NCF& 11;2;2 \\
 \enddata
  \tablerefs{(1) \cite{af98a}, (2) \cite{reese02}, (3) \cite{david1993}, (4) \cite{ettori},
  (5) \cite{vikhlinin}, (6) \cite{schindler}, (7) \cite{allen01}, (8) \cite{peterson}, (9) \cite{hb98},
  (10) \cite{donahue}, (11) \cite{mark96}}
  \tablenotetext{a}{The X-ray emission weighted temperature.} \tablenotetext{b}{The cooling flow corrected X-ray emission weighted
  temperature.}
  \tablenotetext{c}{No confidence intervals were given for these parameters.  It is assumed
  their uncertainty is comparable to the other clusters in our sample.}
\end{deluxetable}

\begin{deluxetable}{lcc}
\tablecaption{Differential Atmospheric Template Factors
\label{tab:alphabeta}}\tablewidth{0pt}
 \tablehead{
 \colhead{Cluster} & \colhead{$\alpha$} & \colhead{$\gamma$} 
  }
 \startdata
    A520 & 0.6785 & -1.4426 \\ 
    A545 & 0.6612 & -1.4493 \\
    A697 & 0.6848 & -1.4374 \\ 
    A773 & 0.6861 & -1.4380 \\ 
    MS1054 & 0.7464 & -1.5391 \\
    RXJ1347 & 0.6703 & -1.3773 \\ 
    A2204 & 0.6819 & -1.4103 \\ 
    A2390 & 0.6676 & -1.4191 \\ 
\enddata
\end{deluxetable}

\begin{deluxetable}{lccc}
\tablecaption{Cluster $\raoff$ and New Multi-Frequency $y_0$
Results \label{tab:finalfit}}\tablewidth{0pt}
 \tablehead{
   \colhead{Cluster} & \colhead{Date} & \colhead{$\Delta$ RA (arcsec)} &
   \colhead{$y_0\times 10^4$}
}
 \startdata
 A520 & Dec02 & $103^{+51}_{-53}$ & $2.00^{+0.70}_{-0.73} $ \\
 A697 & Mar03 & $-25^{+16}_{-17}$ & $4.79^{+1.05}_{-1.06} $\\
 A773 & Mar03 & $7^{+25}_{-27}$ & $4.23^{+2.00}_{-2.32} $ \\
 RXJ1347 & Mar03 & $15^{+12}_{-12}$ & $10.65^{+2.82}_{-2.84} $ \\
 A2204 & Mar03 & $3^{+21}_{-20}$ & $2.53^{+0.77}_{-0.80} $ \\
 A2390 & Dec02 & $-60^{+28}_{-29}$ & $3.61^{+0.73}_{-0.74}$\tablenotemark{a} \\
 \enddata
  \tablenotetext{a}{The central Comptonization of A2390 is calculated at $\raoff=0$ for reasons given in section \ref{sec:centerfit}}
\end{deluxetable}

\begin{deluxetable}{lcccc}
\tablecaption{Cluster
Positions\label{tab:a520center}}\tablewidth{0pt}
  \tablehead{ & & \colhead{R.A.\tablenotemark{a}} & \colhead{Decl.\tablenotemark{a}} & \\
\colhead{Cluster} & \colhead{Instrument} & \colhead{(J2000)} &
\colhead{(J2000)} & \colhead{Ref.}}
 \startdata
A520 & PSPC & 04 54 07.4 & $+$02 55 12.1 & 1 \\
\nodata & HRI & 04 54 10.1 & $+$02 55 27.0 & 2 \\
\nodata & SuZIE & 04 54 $14.3^{+3.4}_{-3.6}$ & \nodata & 3 \\
\nodata  & \nodata  & \nodata  & \nodata & \nodata \\
A2390 & PSPC & 21 53 36.7 & $+$17 41 31.2 & 1 \\
\nodata & HRI & 21 53 36.5 & $+$17 41 45.0 & 2 \\
\nodata & SuZIE & 21 53 $36.4^{+1.1}_{-1.2}$ & \nodata & 4 \\
\nodata & SuZIE & 21 53 $32.9^{+1.7}_{-1.9}$ & \nodata & 3 \\
\enddata
\tablerefs{(1) \citet{1998MNRAS.301..881E}(2) \citet{allen00}
(3) This paper (4) \citet{benson}}
 \tablenotetext{a}{Units of RA are
hours, minutes and seconds and units of declination are degrees,
arcminutes and arcseconds}
\end{deluxetable}

\begin{deluxetable}{lccc}
\tablecaption{Summary of SuZIE Multi-Frequency Central
Comptonization Results \label{tab:clustersummary}}\tablewidth{0pt}
 \tablehead{
   \colhead{Cluster} & \colhead{Date} &
    \colhead{$y_0\times 10^4$} & \colhead{Ref.}}
 \startdata
 A697 & Mar03 & $4.79^{+1.05}_{-1.06} $ & 1 \\
 A773 & Mar03 & $4.23^{+2.00}_{-2.32} $  & 1 \\
 RXJ1347 & Mar03 & $10.65^{+2.82}_{-2.84} $ & 1 \\
 A2204 & Mar03 & $2.53^{+0.77}_{-0.80} $ & 1 \\
 A520 & Dec02 & $2.00^{+0.70}_{-0.73} $ & 1 \\
 A2390 & Nov00/Dec02 & $3.56^{+0.52}_{-0.51}$ & 1,2 \\
 Zw3146 & Nov00 & $3.62^{+1.83}_{-2.52} $ & 2 \\
 A2261 & Mar99 & $7.41^{+1.95}_{-1.98} $ & 2 \\
 MS0451 & Nov96/97/00 & $2.84^{+0.52}_{-0.52} $ & 2 \\
 Cl0016 & Nov96 & $3.27^{+1.45}_{-2.86} $ & 2 \\
 A1835 & Apr96 & $7.66^{+1.64}_{-1.66} $ & 2 \\
 A1689 & Apr94/May94 & $3.43^{+0.59}_{-0.59} $ & 3\\
 A2163 & Apr93/May93 & $3.62^{+0.48}_{-0.48} $ & 3\\
 \enddata
\tablerefs{(1) This paper (2) \citet{benson}(3) \citet{swlh97b}}
\end{deluxetable}


\begin{deluxetable}{lc}
\tablecaption{Summary of SuZIE 145 GHz Central Comptonization
Results \label{tab:d3y0}}\tablewidth{0pt}
 \tablehead{ \colhead{Cluster} &
   \colhead{$y_0\times 10^4$} }
 \startdata
 A697 &  $3.55^{+0.57}_{-0.53} $ \\
 A773 & $3.37^{+0.73}_{-0.66} $ \\
 RXJ1347 & $12.31^{+1.89}_{-1.72}$ \\
 A2204 & $2.44^{+0.43}_{-0.39} $  \\
 A520 & $1.65^{+0.45}_{-0.41} $  \\
 A2390 & $3.57^{+0.42}_{-0.42}$  \\
 Zw3146 & $5.65^{+1.78}_{-1.58} $\\
 A2261 & $6.01^{+0.93}_{-0.81} $  \\
 MS0451 & $3.12^{+0.30}_{-0.29} $ \\
 Cl0016 & $2.31^{+0.93}_{-0.90} $  \\
 A1835 & $6.70^{+1.40}_{-1.24} $  \\
 A1689 & $5.20^{+0.58}_{-0.52} $  \\
 A2163 & $3.25^{+0.40}_{-0.39} $ \\
 A545\tablenotemark{a} & $1.26^{+0.39}_{-0.30}$ \\
 MS1054\tablenotemark{a} &$3.87^{+1.19}_{-1.12}$ \\
 \enddata
\tablenotetext{a}{These clusters were observed by SuZIE and
detected at 145 GHz but lacked the sensitivity at 221 and 355 GHz
to constrain their peculiar velocities.}
\end{deluxetable}

\begin{deluxetable}{lccccc}
\tablecaption{Re-Analysis of SuZIE~I Observations
\label{tab:suziei}}\tablewidth{0pt} \tablehead{
  & \colhead{$kT_e$} & & \colhead{$\theta_{c}$} & & \colhead{IC Gas} \\
  \colhead{Cluster} & \colhead{(keV)} & \colhead{$\beta$} &
  \colhead{(arcsec)} & \colhead{$y_0\times 10^4$} & \colhead{Ref} }
 \startdata
 A1689 &  $9.66$ & $0.609$ & $26.6$ & $5.20^{+0.58}_{-0.52} $ & 1\\
  \nodata &  $8.2$ & $0.78$ & $67.8$ & $3.67^{+0.40}_{-0.38} $ & 2\\
  A2163 &  $12.2$ & $0.674$ & $87.5$ & $3.25^{+0.40}_{-0.39} $ & 1\\
 \nodata &  $12.4$ & $0.616$ & $72.0$ & $3.48^{+0.42}_{-0.42} $ & 2\\
 \enddata
 \tablerefs{(1) \citet{reese02} (2) \citet{swlh97b}}
\end{deluxetable}


\begin{deluxetable}{lcccccc}
\tablecaption{Integrated SZ Flux Results
\label{tab:cluststuff}}\tablewidth{0pt}
 \tablehead{ & & & $d_A$ & \multicolumn{2}{c}{$r_{2500}$} & $S(r_{2500})$\tablenotemark{a} \\
\colhead{Cluster} & $z$ &
   $E(z)$\tablenotemark{b} & \colhead{(MPc)} & \colhead{(kPc)} & \colhead{(arcsec)} & \colhead{(mJy)} }
 \startdata
A697 & 0.282 & 1.154 & 616 & 373 & 125 & $-245^{+45}_{-48}$ \\
A773 & 0.217 & 1.114 & 508 & 399 & 162 & $-326^{+71}_{-78}$ \\
R1347 & 0.451 & 1.271 & 833 & 448 & 111 & $-229^{+35}_{-38}$ \\
A2204 & 0.152 & 1.076 & 382 & 443 & 240 & $-266^{+50}_{-62}$ \\
A520 & 0.199 & 1.103 & 475 & 375 & 163 & $-212^{+55}_{-60}$ \\
A2390 & 0.232 & 1.123 & 534 & 465 & 179 & $-360^{+56}_{-55}$ \\
Zw3146 & 0.291 & 1.160 & 630 & 486 & 159 & $-109^{+37}_{-43}$ \\
A2261 & 0.224 & 1.118 & 520 & 413 & 164 & $-437^{+92}_{-167}$ \\
MS0451 & 0.550 & 1.348 & 926 & 390 & 86.9 & $-73.9^{+8.6}_{-8.6}$ \\
Cl0016 & 0.546 & 1.345 & 923 & 290 & 64.8 & $-47.2^{+19.3}_{-20.0}$ \\
A1835 & 0.252 & 1.135 & 568 & 379 & 138 & $-221^{+42}_{-48}$ \\
A1689 & 0.183 & 1.094 & 444 & 438 & 203 & $-459^{+50}_{-60}$ \\
A2163 & 0.202 & 1.105 & 480 & 465 & 200 & $-533^{+68}_{-74}$ \\
A545 & 0.153 & 1.077 & 384 & 321 & 172 & $-174^{+58}_{-146}$ \\
MS1054 & 0.823 & 1.587 & 1095 & 189 & 35.6 & $-30.3^{+12.8}_{-12.5}$ \\
\enddata
\tablenotetext{a}{The integrated SZ flux, $S(r_{2500})$, is
calculated assuming the SuZIE~II 145 GHz
band.}\tablenotetext{b}{$E(z)^2\equiv \Omega_M (1+z)^3 +
(1-\Omega_M-\Omega_{\Lambda})(1+z)^2 +
\Omega_{\Lambda}$}\tablenotetext{c}{For all calculations where
cosmology is relevant, we assume a standard $\Lambda$CDM cosmology
in a flat universe with $\Omega_M=0.3$, $\Omega_{\Lambda}=0.7$,
and $h=1$.}
\end{deluxetable}

\begin{deluxetable}{ccc}
\tablecaption{XMM Radial Temperature Profile\tablenotemark{a}
\label{tab:tera1835}}\tablewidth{0pt}
 \tablehead{ \colhead{$r_{\rm inner}$} & \colhead{$r_{\rm outer}$} & \colhead{$kT_{e}$} \\ \multicolumn{2}{c}{(')}& \colhead{(keV)} }
 \startdata
0.0 & 0.25 & $4.11\pm0.12$ \\
0.25 & 0.75 & $8.03\pm0.39$ \\
0.75 & 1.5 & $7.12\pm0.36$ \\
1.5 & 2.25 & $7.70^{+0.87}_{-0.77}$ \\
2.25 & 3.33 & $8.55^{+1.36}_{-1.84}$ \\
3.33 & 6.0 & $7.72^{+3.99}_{-2.12}$ \\
\enddata
\tablenotetext{a}{Taken from \citet{majer02}.}
\end{deluxetable}

\begin{deluxetable}{lcccc}
\tablecaption{The Effect of Thermal Structure on the Results for
A1835\label{tab:a1835}}\tablewidth{0pt}
 \tablehead{ & & \colhead{$kT_{e}(r>6')$\tablenotemark{a}} & &
 \colhead{$S(r_{2500})$\tablenotemark{b}} \\ \colhead{Analysis} &
 \colhead{Thermal Structure} & \colhead{(keV)} &
 \colhead{$y_0\times 10^4$} &
 \colhead{(mJy)} }
 \startdata
Multi-frequency & Isothermal & 8.2 & $7.66^{+1.61}_{-1.66}$ & \nodata \\
 & XMM & 7.6 & $7.50^{+1.60}_{-1.61}$ & \nodata \\
 & XMM & 4.11 & $7.46^{+1.58}_{-1.59}$ & \nodata \\
 & Isothermal & 11.5 & $7.99^{+1.70}_{-1.72}$ & \nodata \\
 & Isothermal & 4.11 & $7.19^{+1.56}_{-1.56}$ & \nodata \\
 \nodata & \nodata & \nodata & \nodata & \nodata \\
145 GHz only & Isothermal & 8.2 & $6.70^{+1.40}_{-1.24}$ & $-221^{+42}_{-48}$ \\
 & XMM & 7.6 & $6.63^{+1.46}_{-1.26}$ & $-213^{+41}_{-47}$ \\
 & XMM & 4.11 & $6.71^{+1.46}_{-1.28}$ & $-215^{+41}_{-47}$ \\
 & Isothermal & 11.5 & $6.87^{+1.27}_{-1.19}$ & $-226^{+39}_{-42}$  \\
 & Isothermal & 4.11 & $6.29^{+2.11}_{-1.50}$ & $-204^{+49}_{-68}$  \\
 \enddata
\tablenotetext{a}{The electron temperature at a radius greater
than 6'. For the isothermal cases this temperature is equal to the
temperature assumed throughout the cluster.} \tablenotetext{b}{The
same definition of $r_{2500}$ is used for all models, where we
have assumed $kT_e=8.2$ keV in the calculation of $r_{2500}$.}
\end{deluxetable}

\begin{deluxetable}{ccccccccc}
\tablecaption{Central Comptonization Results with Different Priors
on Peculiar Velocity. \label{tab:y0fits}} \tablewidth{0pt}
  \tablehead{
  $\sigma_{v}$ & \multicolumn{8}{c}{$y_0 \times 10^4$} \\
    {(km s$^{-1}$)} & A697 & A773 & A520 & RXJ1347 & MS0451 &
    Cl0016 & A1835 & A2261 }
\startdata \multicolumn{9}{c}{BIMA/OVRO} \\
      0 & $2.75^{+0.33}_{-0.32} $ & $2.45^{+0.30}_{-0.31} $
   & $1.30^{+0.16}_{-0.20} $ & $7.65^{+0.67}_{-0.67} $ & $2.80^{+0.16}_{-0.21} $
   & $2.40^{+0.19}_{-0.21} $ & $4.85^{+0.31}_{-0.31} $ & $3.30^{+0.37}_{-0.40}
   $\\
   500 & $2.75^{+0.36}_{-0.34} $ & $2.45^{+0.32}_{-0.34} $ & $1.25^{+0.22}_{-0.17} $
  & $7.65^{+0.77}_{-0.76} $ & $2.75^{+0.25}_{-0.19} $ & $2.40^{+0.24}_{-0.25}
  $& $4.85^{+0.41}_{-0.40} $ & $3.25^{+0.45}_{-0.39} $ \\
   2000 & $2.70^{+0.71}_{-0.55} $ & $2.35^{+0.70}_{-0.49} $ & $1.20^{+0.45}_{-0.26}$
  & $7.40^{+2.00}_{-1.41} $ &  $2.70^{+0.62}_{-0.45} $ & $2.25^{+0.84}_{-0.48}
  $& $4.65^{+1.46}_{-0.92} $ & $3.15^{+0.98}_{-0.66} $ \\
  \multicolumn{9}{c}{SuZIE} \\
0 & $3.92^{+0.65}_{-0.64} $ & $4.54^{+1.36}_{-1.36} $ &
$1.84^{+0.59}_{-0.59} $ & $11.70^{+1.85}_{-1.85} $ &
$3.30^{+0.30}_{-0.30} $ &
   $1.87^{+0.86}_{-0.85} $ & $8.00^{+1.14}_{-1.14} $ & $6.24^{+0.85}_{-0.84} $ \\
500 & $4.04^{+0.77}_{-0.72} $ & $4.53^{+1.45}_{-1.40} $  & $1.88^{+0.59}_{-0.59} $  & $11.47^{+2.12}_{-2.00} $
 & $3.22^{+0.35}_{-0.33} $  & $1.88^{+0.91}_{-0.87} $  & $7.97^{+1.22}_{-1.19} $  & $6.33^{+1.04}_{-0.97} $  \\
2000 & $4.61^{+1.03}_{-1.00}$ & $4.43^{+1.85}_{-1.60}$ & $2.01^{+0.68}_{-0.67}$ & $10.87^{+2.58}_{-2.45}$
 & $2.97^{+0.45}_{-0.44}$ & $2.03^{+1.80}_{-1.13}$ & $7.78^{+1.49}_{-1.41}$ & $6.99^{+1.77}_{-1.60}$ \\
\enddata
\end{deluxetable}

\begin{deluxetable}{lccccc}
\tablecaption{Fits to ${\rm log}\left[\frac{S(r_{2500}) d_A^2
E(z)}{\rm Jy Mpc^2}\right] = A + B\, {\rm log}\left[\frac{T_X}{\rm
keV}\right]$ \label{tab:szvtefits}} \tablewidth{0pt}
  \tablehead{\colhead{Sub-sample} &  \colhead{$A$} & \colhead{$B$}
  &\colhead{$N$} &
  \colhead{$\chi^2$} & \colhead{$\chi^2_{\rm red}$\tablenotemark{a}}} \startdata
\multicolumn{6}{c}{Cooling-flow Corrected Temperatures} \\
All & $2.76 \pm 0.41$ & $2.21 \pm 0.41$ & 15 & 6.52 & 0.50 \\
Only NCF & $2.84 \pm 0.72$ & $2.13 \pm 0.71$ & 8 & 2.08 & 0.35 \\
Only CF & $2.78 \pm 0.52$ & $2.25 \pm 0.50$ & 7 & 4.47 & 0.89 \\
\multicolumn{6}{c}{Cooling-flow Un-Corrected Temperatures} \\
All & $2.17 \pm 0.52$ & $2.89 \pm 0.54$ & 15 & 16.3 & 1.25 \\
Only NCF & $2.84 \pm 0.72$ & $2.13 \pm 0.71$ & 8 & 2.08 & 0.35 \\
Only CF & $1.70 \pm 0.63$ & $3.42 \pm 0.67$ & 7 & 9.15 & 1.83 \\
\enddata
\tablenotetext{a}{$\chi^2_{\rm red} = \chi^2 / (N-2)$, where $N$
is the number of clusters in the sub-sample.}
\end{deluxetable}

\begin{deluxetable}{lccccc}
\tablecaption{Fits to ${\rm log}\left[\frac{y_0}{E(z)}\right] = A
+ B\, {\rm log}\left[\frac{T_X}{\rm keV}\right]$
\label{tab:yvtefits}} \tablewidth{0pt}
  \tablehead{\colhead{Sub-sample} &  \colhead{$A$} & \colhead{$B$}
  &\colhead{$N$} &
  \colhead{$\chi^2$} & \colhead{$\chi^2_{\rm red}$\tablenotemark{a}}} \startdata
All & $-2.35 \pm 0.57$ & $2.90 \pm 0.57$ & 15 & 38.0 & 2.93 \\
All(-A1835) & $-2.45 \pm 0.51$ & $2.94 \pm 0.50$ & 14 & 15.0 & 1.25 \\
Only NCF & $-0.34 \pm 0.49$ & $0.77 \pm 0.48$ & 8 & 7.1 & 1.18 \\
Only CF & $3.55 \pm 9.64$ & $-2.88 \pm 2.79$ & 7 & 55.5 & 11.1 \\
Only CF(-A1835) & $-1.11 \pm 12.8$ & $1.52 \pm 1.79$ & 6 & 32.4 & 8.1 \\
\enddata
\tablenotetext{a}{$\chi^2_{\rm red} = \chi^2 / (N-2)$, where $N$
is the number of clusters in the sub-sample.}
\end{deluxetable}

\clearpage

\end{document}